\documentclass[a4paper,11pt]{article}
\pdfoutput=1 

\usepackage{jheppub} 
                     
\usepackage{url}
\usepackage{graphicx}
\usepackage{caption}
\usepackage{amsmath}
\usepackage{multirow}
\usepackage{multirow}
\usepackage{subcaption}
\usepackage{floatrow}
\usepackage[dvipsnames]{xcolor}
\usepackage{float}
\usepackage{amssymb}
\usepackage{hyperref}
\usepackage{color}

\newcommand{\be}{\begin{equation}}
\newcommand{\ee}{\end{equation}}
\newcommand{\ben}{\begin{displaymath}}
\newcommand{\een}{\end{displaymath}}
\newcommand{\bea}{\begin{eqnarray}}
\newcommand{\eea}{\end{eqnarray}}
\newcommand{\rf}[1]{(\ref{#1})}

\def\be{\begin{equation}}
\def\ee{\end{equation}}
\def\bea{\begin{eqnarray}}
\def\eea{\end{eqnarray}}
\def\ba{\begin{array}}
\def\ea{\end{array}}

\def\vp{\varphi}

\def\beq{\begin{equation}}
\def\eeq{\end{equation}}

\newcommand{\Mp}{M_\mathrm{Pl}}

\definecolor{verde}{rgb}{0,0.5,0}

\def\Mp{M_{\rm Pl}}

\begin{document}

\title{\bf \huge{Post-inflationary enhancement of adiabatic perturbations in modular cosmology}}
\author[a,b]{Rodrigo~Gonzalez~Quaglia,}
\author[a]{Martino
Michelotti,}
\author[a]{Diederik Roest,}
\author[c,d]{\hskip 3cm John Joseph Carrasco,}
\author[e]{Renata Kallosh,}
\author[e]{ Andrei Linde}

\affiliation[a]{Van Swinderen Institute for Particle Physics and Gravity,
University of Groningen, Nijenborgh 3, 9747 AG Groningen, The Netherlands}
\affiliation[b]{Instituto de Ciencias F\'{\i}sicas, Universidad Nacional
Aut\'onoma de M\'exico,\\ Av. Universidad s/n, Cuernavaca, Morelos, 62210, Mexico}
\affiliation[c]{Amplitudes and Insights Group, Department of Physics and Astronomy,
Northwestern University, Evanston, IL 60208, USA}
\affiliation[d]{Center for Interdisciplinary Exploration and Research in Astrophysics (CIERA)
Northwestern University, 1800 Sherman Ave, Evanston, IL 60201,USA}
\affiliation[e]{Stanford Institute for Theoretical Physics and Department of Physics,
Stanford University, Stanford, CA 94305, USA}

\emailAdd{r.gonzalez.quaglia@rug.nl}\emailAdd{m.michelotti@rug.nl}\emailAdd{d.roest@rug.nl}\emailAdd{ carrasco@northwestern.edu}\emailAdd{kallosh@stanford.edu}
\emailAdd{alinde@stanford.edu}

\abstract{We show that multi-field inflationary models with negligible turning in field space during inflation can lead to an effective sourcing of adiabatic from entropic perturbations {\it after} the end of inflation. We illustrate this general phenomenon with a detailed analysis of an inflationary model whose scalar potential is determined by modular invariance. Its entropic perturbations are frozen during inflation, but instead, they are converted into adiabatic perturbations in the first post-inflationary $e$-folds. The curvature power spectrum, giving rise to CMB fluctuations, reaches a novel and enhanced plateau in this process; we address the implications for the inflationary observables $A_{s}$, $n_{s}$ and $r$.}

\maketitle
\flushbottom

\parskip 5pt
\section{Introduction}\label{Introduction}

The inflationary paradigm has become the most widely accepted description of the early universe \cite{Guth:1980zm,Linde:1984ir,Lyth:1998xn}. It poses a period of accelerated expansion in the very early universe, responsible not only for solving the $\Lambda$CDM puzzles of cosmology but also for providing the origin of the seeds needed for structure formation in the universe. The simplest model of inflation is characterized by a single scalar field with a potential possessing a sufficiently flat region, see \textit{e.g.}~\cite{Linde:1983gd,Linde:1993cn}.

Moreover, it has been found that the issue of quantum corrections affecting the flatness of the potential must be addressed in order to build a compelling inflationary model \cite{Berera:2003yyp,Easson:2009kk,Ashoorioon:2011aa}. In this regard, a typical approach to solving this problem is to consider a symmetry in the model that suppresses higher-order interaction terms. Indeed, this idea leads to the widely studied natural inflation model \cite{Freese:1990rb,Adams:1992bn} which utilizes a discrete shift symmetry to protect the flatness of the potential.

In this paper, following the recent flourishing of modular cosmology \cite{Kallosh:2024ymt,Kallosh:2024pat,Kallosh:2024whb,Ding:2024euc,Aoki:2024ixq,Carrasco:2025rud,Aoki:2025wld} as an active and promising research direction, we aim to continue the study of these types of models characterized by the invariance under $SL(2,{\mathbb Z})$ transformations, with the intention of numerically confirming and extending previous results, with a specific focus on those in Ref.~\cite{Kallosh:2024whb}.

The generic Lagrangian describing modular models builds on that of $\alpha$-attractors \cite{Kallosh:2013hoa,Roest:2013fha,Ferrara:2013rsa,Kallosh:2013maa,Kallosh:2013daa,Kallosh:2013yoa,Kallosh:2013tua,Cecotti:2014ipa,Kallosh:2014rga,Kallosh:2014laa,Galante:2014ifa,Carrasco:2015pla,Carrasco:2015rva,Carrasco:2015iij,Kallosh:2016sej,Iacconi:2024hmg} and has the form
\begin{equation}\label{modularlagrangian}
    \frac{{\cal L}(\tau,\bar{\tau})}{\sqrt{-g}}=\frac{\Mp^2}{2}R-\frac{3\alpha}{4}\frac{\partial\tau\partial\bar{\tau}}{({\rm Im}\tau)^2}-V(\tau,\bar{\tau})\,,
\end{equation}
where $R$ is the Ricci scalar, $g$ is the determinant of the metric, $\tau$ is a complex scalar field living in the upper half-plane, {\it i.e.} ${\rm Im}\tau\geq0$ and $\alpha$ is proportional to the inverse of the Kahler curvature
 \footnote{Lower bounds on the parameter $\alpha$ have been studied in \cite{German:2022sjd,Iacconi:2023mnw}.} of the $SL(2,\mathbb R)/U(1)$ coset space \cite{Kallosh:2013yoa,Carrasco:2015uma,Kallosh:2025jsb}.

Due to the isometries of the hyperbolic space, the kinetic term is manifestly invariant under the $SL(2,{\mathbb R})$ transformation
\begin{equation}\label{modulartransformation}
    \tau\xrightarrow[]{}\frac{a\tau+b}{c\tau+d}, \quad \mbox{with} \quad ad-cb=1\,,
\end{equation}
for any real constants $a,b,c,d$. Introducing hyperbolic coordinates $\tau=\theta+ie^{\sqrt{\frac{2}{3\alpha}}\varphi}$, the kinetic term is then written as
\begin{equation}\label{ModularKineticterm}
    -\frac{3\alpha}{4}\frac{\partial\tau\partial\bar{\tau}}{({\rm Im}\tau)^2}=-\frac{1}{2}(\partial\varphi)^2-\frac{3\alpha}{4}e^{-2\sqrt{\frac{2}{3\alpha}}\varphi}(\partial\theta)^2.
\end{equation}
In this form, it is apparent that the modular model field-space metric is non-trivial, which leads to interesting multi-field dynamics. Indeed, this hyperbolic field metric has been employed in diverse contexts such as the study of $T$ and $E$-models. In Ref.~\cite{Iarygina:2020dwe}, it was found that, for a wide range of initial conditions, these two exhibit identical multi-field dynamics during inflation characterized by two approximately single-field stages connected
by a sharp turn in field space. Another interesting scenario is the one studied in Ref.~\cite{Achucarro:2017ing}, where both the dilaton and the inflaton field are light. In this reference, unlike the case of sharp turning in field space \cite{Iarygina:2020dwe}, the model's dynamics exhibits a constant turning, which makes the curvature power spectrum coincide with the single-field estimate.

As we shall see, the turning in field space is a crucial quantity, as it controls the efficiency of sourcing adiabatic information from entropic perturbations. In the present paper, we will build upon the work in Ref.~\cite{Kallosh:2024whb}, where this turning is found to be negligible during inflation, but potentially significant immediately after the end of inflation.  Our main point here is to extend the investigation of perturbations beyond the end of inflation, where the multi-field dynamics becomes important, and the sourcing of perturbations is an interesting question to address.

The paper is structured as follows. In Section \ref{Inflation in modular cosmology} we present and discuss the features of the two models considered in this work and solve for their background evolution. In doing so, we adopt the adiabatic-entropic decomposition as the natural formalism for multi-field models of inflation. In Section \ref{The adiabatic and entropic decomposition}, we move on to studying the multi-field system of perturbations during inflation, focusing on the calculation of the power spectra and showing in particular how adiabatic and entropic perturbations are coupled and source each other. Section \ref{Postinflationary evolution} extends this analysis to the post-inflationary period, where we demonstrate how an efficient sourcing of perturbations may occur after the end of inflation. In that section, we also check whether the process of reheating may affect our analysis. Finally, Section \ref{Discussion and conclusions} includes our concluding remarks.   Appendix A includes details on a modular invariant model based on a different $SL(2,\mathbb{Z})$-invariant, the $j$-function.


\section{Inflation in modular cosmology}\label{Inflation in modular cosmology}
Along the modular invariant kinetic term Eq.~\eqref{ModularKineticterm}, we need to also introduce a modular potential $V(\tau,\bar{\tau})$ which breaks the continuous $SL(2,{\mathbb R})$ symmetry down to $SL(2,{\mathbb Z})$, {\it i.e.} $(a,b,c,d)\in \mathbb Z$. This is completely analogous to the case of natural inflation, where the continuous shift symmetry in the kinetic term is broken to its discrete version whenever the sinusoidal potential is included.

In order to explore the inflationary dynamics of the multi-field system, we shall consider two models: the first respects the $SL(2,\mathbb Z)$ invariance of modular cosmology, while the second is described by an \textit{ad-hoc} potential, crafted to have similar features for it to serve as a simpler toy model.

\subsection{Modular potential}

As mentioned, we are interested in a modular potential which breaks down the $SL(2,\mathbb{R})$ symmetry to $SL(2,\mathbb{Z})$. In this work, we consider an explicit expression in terms of the \textit{Dedekind eta} function\footnote{The previous works Ref.~\cite{Kallosh:2024whb} featured instead the $j$-function, whose fluctuation analysis, however, is somewhat computationally heavier; see Appendix A for the analogous results.},  $\eta_D$, given by
\begin{equation}\label{dedekindmodel}
    V_{\rm mod}(\varphi,\theta)=V_{0}\left(\frac{1-g_0/g(\varphi,\theta)}{1+g_0/g(\varphi,\theta)}\right)\,,
\end{equation}
where $V_0$ is the scale of the potential, and the field dependence is given by the function
\begin{equation}\label{fg}
    g(\varphi,\theta)\equiv \log \left[ e^{\sqrt{\frac{2}{3\alpha}}\varphi}\, \eta_{D}^2 \left(ie^{\sqrt{\frac{2}{3\alpha}}\varphi}-\theta\right) \eta_{D}^2 \left(ie^{\sqrt{\frac{2}{3\alpha}}\varphi}+\theta\right) \right] \,.
\end{equation}
The constant $g_0$ is simply this function evaluated at $g_0 \equiv g\left(\sqrt{\frac{3\alpha}{2}}\log\left[\frac{\sqrt{3}}{2}\right],\frac{1}{2}\right)$, where the potential has a Minkowski minimum (note that $g_0 <0$). The \textit{Dedekind eta} function is defined, for any complex number $\tau=x+iy$ with $y>0$, as 
\begin{equation}
    \eta_{D}(\tau)\equiv e^{\frac{i\pi \tau}{12}}\prod_{n=1}^{\infty}\left(1-e^{2i\pi n\tau}\right)\,,
\end{equation}
which is manifestly invariant under the modular transformation in Eq.~\eqref{modulartransformation}. Moreover, it possesses the asymptotic limit
\begin{equation}
    \lim_{y\rightarrow\infty}\eta_{D}(\tau)\sim e^{-\frac{\pi y}{12}}e^{\frac{i\pi x}{12}}\,.
\end{equation}
Notice that the argument of the logarithm in $g(\varphi,\theta)$ is strictly real due to the property $\bar{\eta}_D(\tau)=\eta_D(-\bar{\tau})$.

We are interested in the asymptotic behaviour of the potential for large $\varphi$ as this is the region where inflation takes place. Keeping the leading dependencies in both $\varphi$ and $\theta$ we obtain
\begin{equation}\label{largephipot}
\begin{aligned}
    \lim_{\varphi\rightarrow\infty}V_{\rm mod}(\varphi,\theta)\sim V_0\Bigg(&1+\frac{6}{\pi} g_0 e^{-\sqrt{\frac{2}{3\alpha}}\varphi}+\frac{18}{\pi^2} g_0\sqrt{\frac{2}{3\alpha}}\varphi e^{-2\sqrt{\frac{2}{3\alpha}}\varphi}+\frac{18}{\pi^2} g_0^2 e^{-2\sqrt{\frac{2}{3\alpha}}\varphi} \\
    &-\frac{72}{\pi^2}g_0e^{-2\sqrt{\frac{2}{3\alpha}}\varphi}e^{-2\pi e^{\sqrt{\frac{2}{3\alpha}}\varphi}}\cos(2\pi\theta)\Bigg)\,.
\end{aligned}
\end{equation}
We see that the construction of the $SL(2,\mathbb Z)$-invariant model is such that the $\theta$ direction is double exponentially suppressed for large $\varphi$ (as discussed in detail in Ref.~\cite{Kallosh:2024whb} for another type of modular invariant), so that the dependence on $\theta$ can be neglected and a single-field $\alpha$-attractor behaviour is recovered. Assuming that no significant dynamics in the $\theta$ direction develops until the end of inflation, it is then possible to derive an analytical solution for the effectively single-field dynamics of $\varphi$. As we will demonstrate, this is precisely what happens when the full model is solved using numerical techniques.

Introducing the number of $e$-folds $N$ as\footnote{Notice that we are defining $N$ so that it remains {\it negative} during inflation. The end of inflation at $N=0$ is conventionally set to correspond to $\epsilon=1$.}
\begin{equation}
    N\equiv\int^t_{t_{\rm end}}Hdt\simeq-\int^\varphi_{\varphi_{\rm end}}\frac{V}{V_\varphi}d\varphi\,,
\end{equation}
where in the second step we used the fact that in the single-field slow-roll approximation $\dot{\varphi}=-\sqrt{2\epsilon}H\simeq V_\varphi H/V$. This integral can be computed explicitly considering the leading $\varphi$ dependence in the asymptotic expansion shown in Eq.~\eqref{largephipot}. To leading order in $\varphi$ and after inverting we obtain
\begin{equation} \label{phiN}
    \varphi(N)\simeq\sqrt{\frac{3\alpha}{2}}\log\left(\frac{4 g_0}{\pi\alpha}N\right)\,,
\end{equation}
where we neglected the contribution of  $\varphi_{\rm end}$ which is a valid assumption early enough during inflation. This result can be used to compute the predictions of the modular-invariant model (Eq.~\eqref{dedekindmodel}) and check that it indeed belongs to the $\alpha$-attractor family. Assuming single-field dynamics, the amplitude of the primordial scalar power spectrum is given by
\begin{equation}\label{AsModpot}
    A_s=\frac{V^3}{12\pi^2V_\varphi^2}\simeq\frac{V_0N_{\rm hc}^2}{18\pi^2\alpha}\,,
\end{equation}
where $N_{\rm hc}\equiv\ln(k/H)$ is evaluated at the time of horizon-crossing of the relevant observable scales, typically set at $N_{\rm hc}=60$ for CMB measurements. The scalar spectral index $n_s$ and the tensor-to-scalar ratio $r$ are then obtained as
\begin{equation}\label{nsrmodpot}
    n_s-1\equiv\frac{d\log A_s}{dN_{\rm hc}}\simeq-\frac{2}{N_{\rm hc}}\,, \quad r\equiv\frac{A_t}{A_s}\simeq\frac{12\alpha}{N_{\rm hc}^2}\,,
\end{equation}
where
\begin{equation}
    A_t\simeq\frac{2V_0}{3\pi^2}\,,
\end{equation}
is the amplitude of the primordial tensor power spectrum.

As expected, these results are the familiar predictions of the $\alpha$-attractor class of models, confirming the behaviour of the $SL(2,\mathbb Z)$-invariant model when the dynamics of $\theta$ can be neglected. This very same conclusion was already obtained in Ref.~\cite{Kallosh:2024whb} for a different model based on the $SL(2,\mathbb Z)$-invariant known as the \textit{Klein $j$-invariant}, which we discuss in Appendix \ref{J function}. The scope of the present work is to check these estimates with full numerical calculations, and moreover to extend these beyond the end of inflation -- where the dynamics of $\theta$ may no longer be negligible and multi-field dynamics may also alter the $\alpha$-attractor predictions. \\

\subsection{Hyperbolic potential}
Before moving on to the study of the modular-invariant model, we will also introduce a second model to facilitate comparison and contrast with our results. Moreover, this second model will serve as an illustration that our results are more general and hold not only for modular-invariant models.

Taking inspiration from the features of the modular potential Eq.~\eqref{dedekindmodel}, we further construct a simpler toy model in terms of the hyperbolic tangent as
\begin{equation}\label{tanhmodel}
    V_{\rm hyper}(\varphi,\theta)=V_{0}\left[\tanh^{2}\left(\frac{\varphi}{\sqrt{6\alpha}}\right)+\beta^{2}e^{-2\sqrt{\frac{2}{3\alpha}}\varphi}e^{-\frac{4}{3\alpha}\varphi^{2}}\cos^{2}(\pi\theta)\right]\,.
\end{equation}
This potential should be regarded as a toy model whose quadratic minimum, in the $\varphi$ direction, is tailored such that it is qualitatively similar to the one found in the modular invariant model Eq.~\eqref{dedekindmodel}. Moreover, requiring the mass matrix, defined as the Hessian evaluated at the minimum of the potential, to be proportional to the identity sets the free parameter $\beta\simeq0.2$.

A second property of this potential is such that it is constructed in order to resemble the suppression along the $\theta$ direction of the modular potential, as can be seen by considering the large $\varphi$ limit of Eq.~\eqref{tanhmodel}
\begin{equation} \label{largephipothyp}
    \lim_{\varphi\rightarrow\infty}V_{\rm hyper}(\varphi,\theta)\sim V_0\left[1-4e^{-\sqrt{\frac{2}{3\alpha}}\varphi}+8e^{-2\sqrt{\frac{2}{3\alpha}}\varphi}+\beta^2e^{-2\sqrt{\frac{2}{3\alpha}}\varphi}\cos^2(\pi\theta)\right]\,,
\end{equation}
which is similar yet not identical to the one in Eq.~\eqref{largephipot}\footnote{Note that this difference can be resolved by a shift in $\varphi$ and hence does not constitute a fundamental distinction.}. Indeed, the leading $\varphi$ dependence is again of the $\alpha$-attractor class, so that the predictions are precisely the same obtained for the modular case, provided the $\theta$ dynamics (which is not doubly exponentially suppressed, but comes in at a subleading order in the exponential expansion) is negligible until the end of inflation. We will verify this explicitly by also solving the background evolution numerically for the hyperbolic potential, Eq.~\eqref{tanhmodel}.

\begin{figure}[t]
       \begin{subfigure}[b]{0.45\textwidth}
        \centering
        \includegraphics[width=1\linewidth]{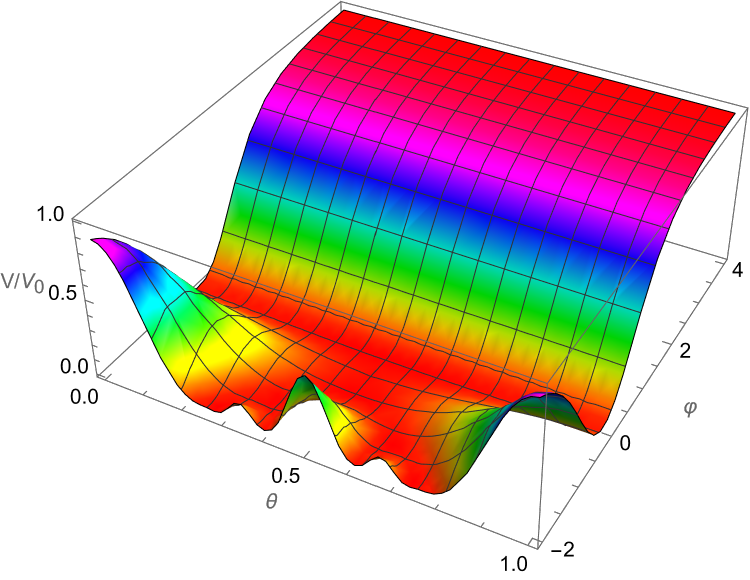}
    \end{subfigure}
    \hfill
    \begin{subfigure}[b]{0.45\textwidth}
        \centering
        \includegraphics[width=1\linewidth]{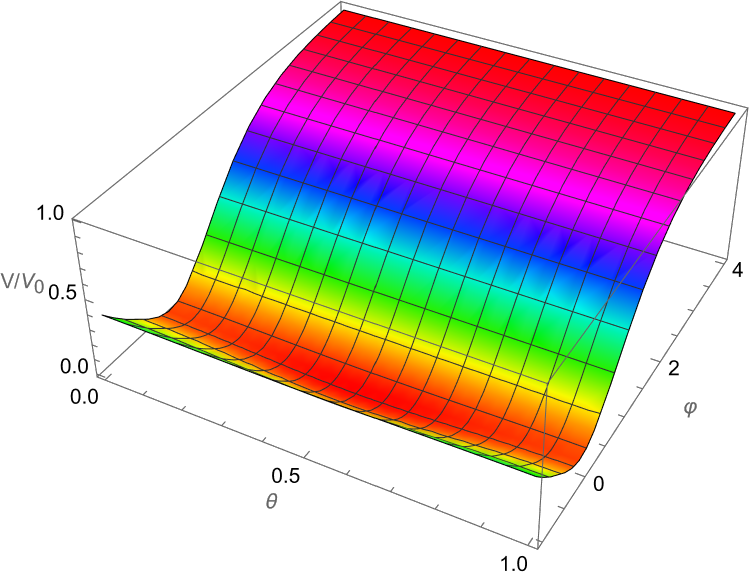} 
    \end{subfigure}
    \caption{\footnotesize {Plot of the modular-invariant potential Eq.~\eqref{dedekindmodel} (left panel) and the hyperbolic potential Eq.~\eqref{tanhmodel} (right panel). Both potentials exhibit a similar flat region for large $\varphi$, where inflation takes place, together with a quadratic-shaped minimum. Here we can appreciate how the hyperbolic model is a simplified version of the modular one, as the latter has a richer structure of vacua, mandated by modular invariance \cite{Kallosh:2024pat}}.}
    \label{Potentials}
\end{figure}

For visual aid, we plot both the modular and hyperbolic potentials in Fig. \ref{Potentials}.

\subsection{The adiabatic-entropic decomposition}

It is convenient to introduce a more natural basis to parametrize the multi-field background trajectories. The idea is to define, at each point in field space, the tangential (\textit{adiabatic}) and orthogonal (\textit{entropic}) direction with respect to the trajectory, defining the so-called adiabatic-entropic decomposition \cite{Wands:2007bd}.

In general, the Lagrangian for a multi-field model of inflation is given by 
\begin{equation}\label{MultifieldAction}
    \frac{{\cal L}(\phi^I)}{\sqrt{-g}}=\frac{1}{2}R-\frac{1}{2}g^{\mu\nu}G_{IJ}(\phi)\partial_{\mu}\phi^{I}\partial_{\nu}\phi^{J}-V(\phi)\,, 
\end{equation}
with $G_{IJ}(\phi)$ the (internal) field-space metric of the set of $n$ scalar fields $\phi^{I}$.
The equations of motion, in a flat FLRW background, derived from this Lagrangian, read 
\begin{equation}
    3H^{2}=\frac{1}{2}\dot{\sigma}^{2}+V(\phi)\,, \quad D_{t}\dot{\phi^{I}}+3H\dot{\phi^{I}}+G^{IJ}\frac{\partial V}{\partial\phi^{J}}=0\,,
\end{equation}
where dots represent derivatives with respect to the cosmic time $t$, and we introduced the generalized kinetic energy and covariant time derivative as
\begin{equation}
    \dot{\sigma}^2\equiv G_{IJ}\dot{\phi}^I\dot{\phi}^J\,, \qquad
    D_{t}\dot{\phi^{I}}\equiv\ddot{\phi^{I}}+\Gamma^{I}_{\ JK}\dot{\phi^{J}}\dot{\phi^{K}}\,.
\end{equation}
The slow-roll parameters in multi-field inflation are then computed as
\begin{equation}
    \epsilon_{H}\equiv -\frac{\dot{H}}{H^{2}}=\frac{\dot{\sigma^{2}}}{2H^{2}}\,,\quad \eta_{H}\equiv\frac{\dot{\epsilon}}{H\epsilon}\,.
\end{equation}
Within this multi-field formalism, the adiabatic-entropic decomposition is introduced as the basis for the tangent field space, in which the metric is locally flat at any point along the background trajectory. This is achieved by introducing the adiabatic unit vector in the tangential direction
\begin{equation}
    e_\sigma^I\equiv\frac{\dot{\phi}^I}{\dot{\sigma}}\,,
\end{equation}
which, together with the entropic unit vector(s), form a complete set of $n$-\textit{beins} of the field space, such that
\begin{equation}
    G_{IJ}e_i^Ie_j^J=\delta_{ij}\,.
\end{equation}
Notice that we adopt uppercase Latin letters $I,J,\ldots$ for curved indices, related to the original field metric in terms of $\varphi$ and $\theta$, and lowercase Latin letters $i,j,\ldots$ for flat indices, related to the flat metric defined by the $n$-beins.

In the two-field case under scrutiny, we have a set of \textit{zweibeins} with a single orthogonal direction, which we identify with the $s$ index. In particular, the orthogonal unit vector can be computed as
\begin{equation}
   e_s^I=-G^{-1/2}\epsilon^{IJ}G_{JK}e^K_\sigma\,,
\end{equation}
where $G$ is the field-metric determinant and $\epsilon^{IJ}$ the totally antisymmetric Levi-Civita symbol. Moreover, the tangential and orthogonal directions are dynamical and subject to change if the background trajectory experiences a turning in field space, as described by the following system of equations
\begin{equation}
    D_te_\sigma^I=H\eta_\perp e_s^I\,, \quad  D_te_s^I=-H\eta_\perp e_\sigma^I\,,
\end{equation}
where we introduced the \textit{turning rate} $\eta_\perp$, which quantifies the amount of bending the field-space trajectory is subject to. Once the zweibeins are known, the turning rate is obtained by inverting the previous relation
 \begin{equation}\label{DefTurnRate}
    \eta_{\perp}=\frac{1}{H}e_{Is}D_{t}e^{I}_{\sigma}=-\frac{1}{H}e_{I\sigma}D_{t}e^{I}_{s}\,.
\end{equation}
For the hyperbolic geometry in Eq.~\eqref{ModularKineticterm}, we have the following explicit zweinbeins components
\begin{equation}
    e^I_\sigma=\frac{1}{\dot{\sigma}}\begin{pmatrix}
        \dot{\varphi} \\
        \dot{\theta}
    \end{pmatrix}\,, \quad e^I_s=\frac{1}{\dot{\sigma}}\begin{pmatrix}
        \sqrt{\frac{3\alpha}{2}}e^{-\sqrt{\frac{2}{3\alpha}}\varphi}\dot{\theta} \\
        \sqrt{\frac{2}{3\alpha}}e^{\sqrt{\frac{2}{3\alpha}}\varphi}\dot{\varphi}
    \end{pmatrix}\,,
\end{equation}
so that the turning rate is given by 
\begin{equation}\label{exactturn}
    \eta_\perp=\sqrt{\frac{2}{3 \alpha }}  \frac{e^{\sqrt{\frac{2}{3\alpha}} \varphi}}{\dot{\sigma}^2 H}\left(V_\theta\dot{\varphi}-\frac{3}{2}\alpha e^{-2\sqrt{\frac{2}{3 \alpha }}\varphi}V_\varphi \dot{\theta}\right)\,.
\end{equation}
We will explicitly evaluate this quantity after solving the background dynamics for both models.

\subsection{Background evolution}

The general background equations of motion derived from the hyperbolic geometry Eq.~\eqref{ModularKineticterm} read
\begin{equation}\label{Friedmanlike}
    3H^2=V(\varphi,\theta )+\frac{3}{4}\alpha \ e^{-2 \sqrt{\frac{2}{3\alpha} } \varphi} \dot{\theta}^2+\frac{1}{2} \dot{\varphi}^2\,,
\end{equation}
\begin{equation}\label{phiEOM}
    \ddot{\varphi}+3H\dot{\varphi}+\frac{\partial V(\varphi,\theta)}{\partial \varphi}+\sqrt{\frac{3\alpha}{2}}e^{-2 \sqrt{\frac{2}{3\alpha}}\varphi} \dot{\theta }^2=0\,,
\end{equation}
\begin{equation}\label{ThetaEOM}
    \ddot{\theta}+3H\dot{\theta}-\sqrt{\frac{2}{3\alpha}}\dot{\theta}\dot{\varphi}+\frac{2e^{2 \sqrt{\frac{2}{3\alpha}} \varphi}} {{3 \alpha }} \frac{\partial V(\varphi,\theta)}{\partial \theta}=0\,.
\end{equation}
This system of equations has been analytically studied in Ref.~\cite{Kallosh:2024whb} for the generic type of potentials we are considering in this work (\textit{i.e.} for potentials exhibiting a strong suppression on the $\theta$ direction). In this paper instead we numerically study the dynamics not only in the $SL(2,{\mathbb Z})$ class of models but also in the toy model with the hyperbolic potential, extending the treatment beyond the end of inflation in order to fully capture the turning in field space and estimate its effect on the inflationary predictions.

\begin{figure}[t!]
    \centering
    \begin{subfigure}[b]{0.45\textwidth}
        \includegraphics[width=1\linewidth]{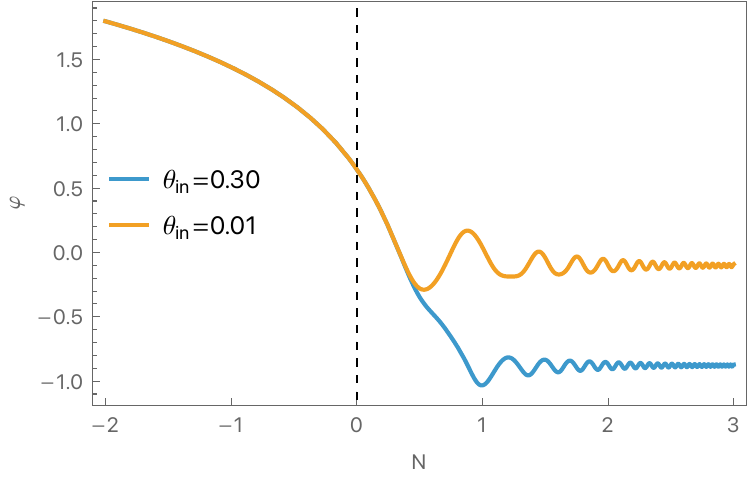}
    \end{subfigure}
    \hfill
    \begin{subfigure}[b]{0.45\textwidth}
        \includegraphics[width=1\linewidth]{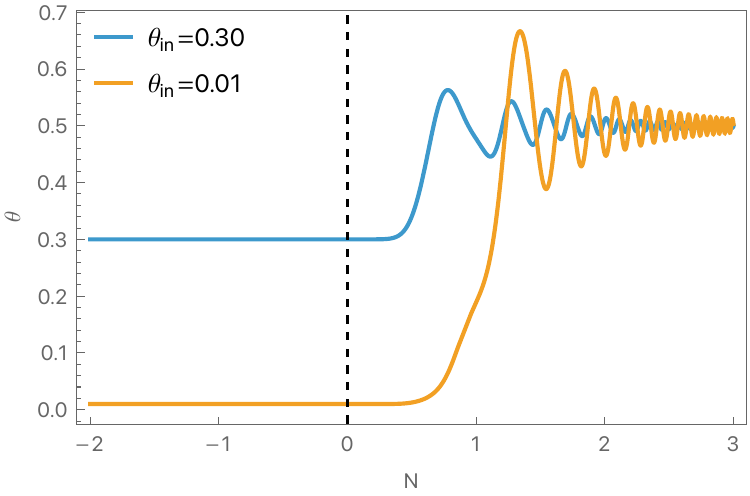}
    \end{subfigure}

    \vfill
    \begin{subfigure}[b]{0.6\textwidth}
        \centering
        \includegraphics[width=0.7\linewidth]{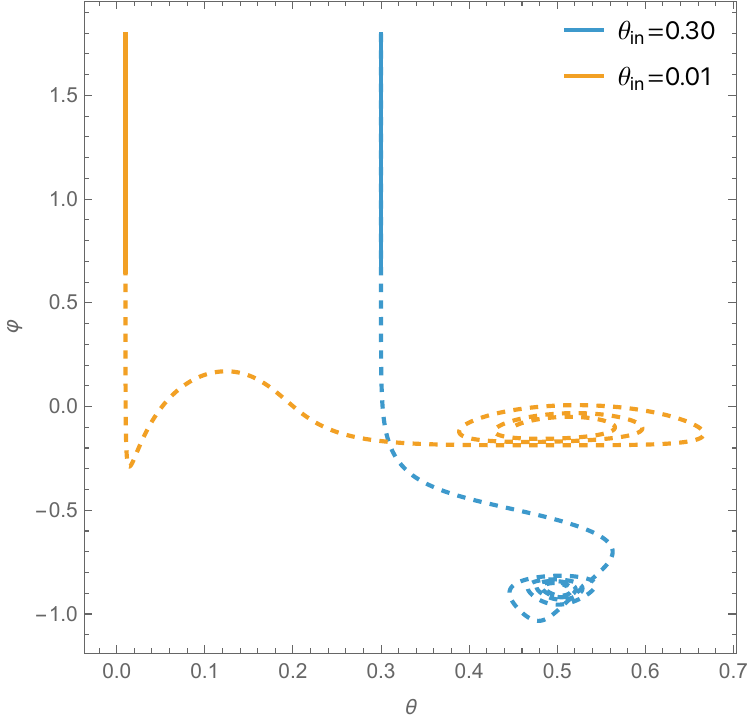}
    \end{subfigure}

    \caption{\footnotesize {Plots depicting the background trajectories for the modular model Eq.~\eqref{dedekindmodel} for $3\alpha=1$ and for two choices of initial angle, $\theta_{\rm in}=0.3$ in blue and $\theta_{\rm in}=0.01$ in yellow. In the upper panels, we report the trajectories for both fields $\varphi$ and $\theta$ for our choices of initial conditions, explicitly showing how different initial angles lead to different minima in the $\varphi$ direction. In both these plots, we denote the end of inflation with a black dashed vertical line at $N=0$. In the bottom panel, we show the parametric trajectories in field space for both initial angles. Here, solid lines represent the inflationary evolution while dashed lines correspond to the post-inflationary dynamics. These trajectories suggest that the modular invariant model Eq.~\eqref{dedekindmodel} experiences geodesic motion during inflation as the $\theta$ field is fixed during this period, with a non-negligible turning afterwards.}}
    \label{dedekindbackgroundplots}
\end{figure}

\begin{figure}[t!]
    \begin{subfigure}[b]{0.45\textwidth}
        \centering
        \includegraphics[width=1\linewidth]{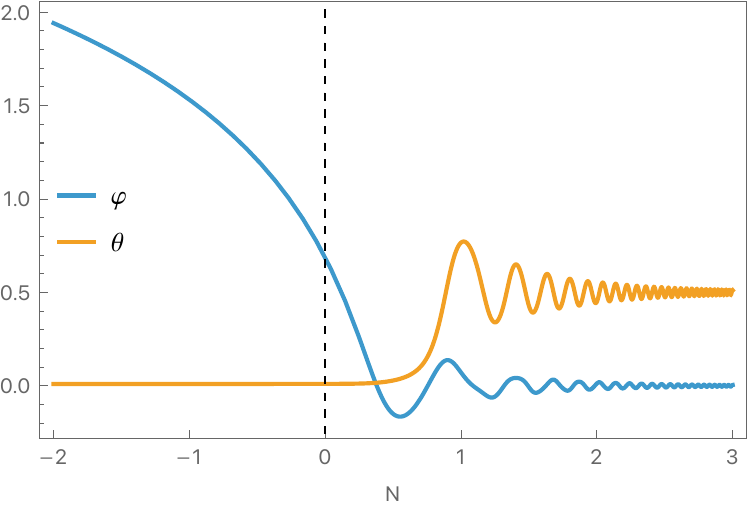}  
        \label{fig:plot2}
    \end{subfigure}
    \hfill
    \begin{subfigure}[b]{0.45\textwidth}
        \centering
        \includegraphics[width=1\linewidth]{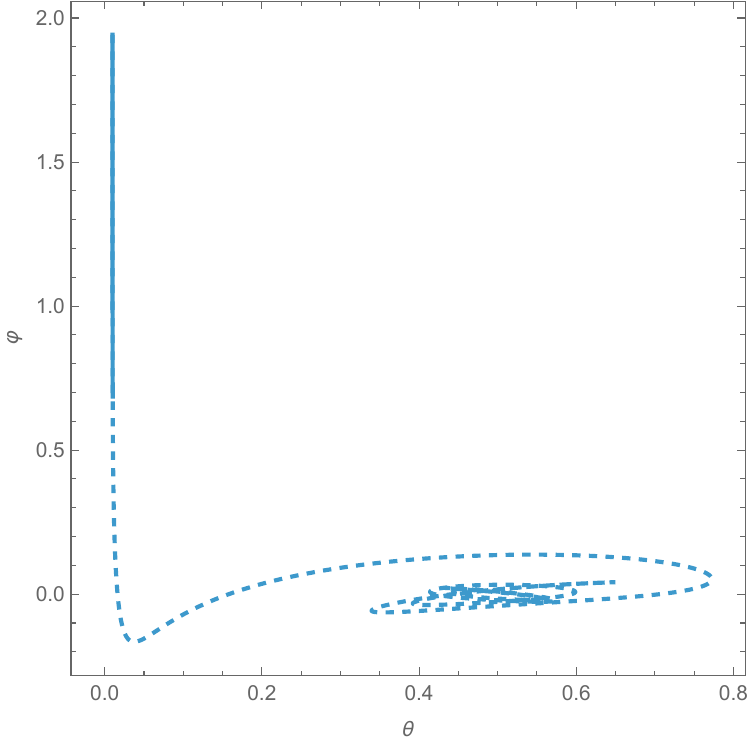} 
        \label{fig:plot4}
    \end{subfigure}

    \caption{\footnotesize {Plots showing the background trajectories for the hyperbolic model  Eq.~\eqref{tanhmodel} setting $\beta=1/5$,  $3\alpha=1$ and $\theta_{\rm in}=0.01$. On the left panel, we plot the background evolution of both $\varphi$ and $\theta$, indicating the end of inflation $(N=0)$ with the vertical dotted line. On the right panel, instead, we show the parametric evolution of both fields, where the solid line represents the evolution during inflation, while the dotted one indicates the post-inflationary behavior of both fields. Both plots show that, during inflation, the $\theta$ field remains essentially constant; thus, during this period, the model undergoes geodesic motion. However, shortly after the end of inflation, the trajectory in field space starts to drastically turn.}}
    \label{tanhbackground}
\end{figure}

The background trajectories for both potentials are depicted in Figures~\ref{dedekindbackgroundplots} and  \ref{tanhbackground} respectively. These numerical solutions confirm our hypothesis that, during inflation, $\theta$ is fixed due to the strong suppression and thus the motion in field space is along geodesics with constant $\theta$. Then, soon after inflation, the field space trajectory turns, leading to potentially richer multi-field dynamics.

For the hyperbolic potential Eq.~\eqref{tanhmodel}, the qualitative picture is similar to the modular-invariant one. Even if this potential does not exhibit such a strong, double exponential, suppression it is still flat enough so that $\theta$ is fixed during inflation. After this, the turning rate starts to increase as $\theta$ looks for its minima (see Fig.  \ref{tanhbackground}), leading to an efficient sourcing of perturbations.

\begin{figure}[t!]
    \begin{subfigure}[b]{0.45\textwidth}
        \centering
        \includegraphics[width=1\linewidth]{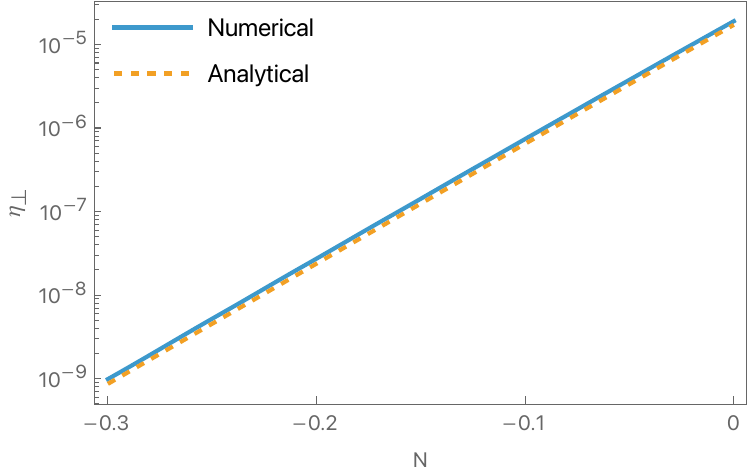}  
        \label{fig:plot2a}
    \end{subfigure}
    \hfill
    \begin{subfigure}[b]{0.45\textwidth}
        \centering
        \includegraphics[width=1\linewidth]{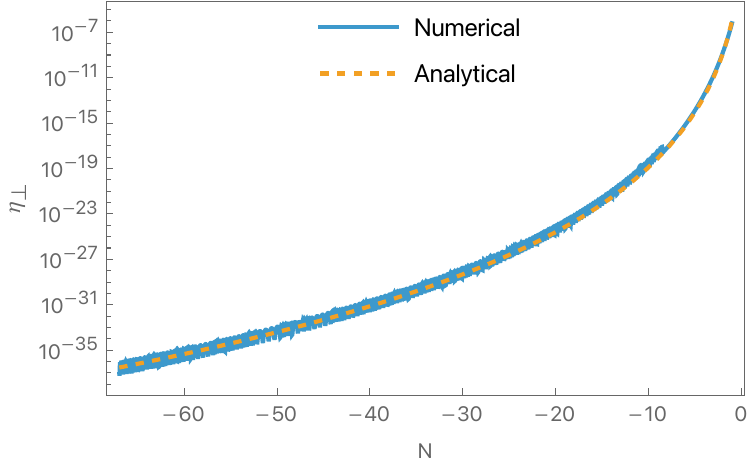} 
        \label{fig:plot4a}
    \end{subfigure}

    \caption{\footnotesize {Plots of the turning rate $\eta_{\perp}$  for both the modular Eq.~\eqref{dedekindmodel} (left panel) and hyperbolic Eq.~\eqref{tanhmodel} (right panel) models. In these plots we are showing the exact numerical solution for the turning rate in Eq.~\eqref{DefTurnRate} in blue and comparing it to the analytical ones (see Eq.~\eqref{turnDEDE}), in dashed yellow, obtained after assuming slow-roll dynamics on $\varphi$ and neglecting contributions from $\dot{\theta}$. We find that these two are in good agreement and that both depict an extremely small turning rate, up to the end of inflation ($N=0$). Note that for numerical accuracy, we only show the last $0.3$ $e$-folds of inflation in the case of the modular invariant model (left panel). However, we expect that the analytical formula (first equation in Eq.~\eqref{DefTurnRate}) holds for the earlier parts of inflation as well.}}
    \label{turns}
\end{figure}

In the slow-roll approximation, we can estimate the turning rate to be 
\begin{equation}\label{approxturn}
    \eta_{\perp}\approx-\frac{\sqrt{6}\ V_{,\theta}}{\sqrt{\alpha}\  V_{,\varphi}}e^{\sqrt{\frac{2}{3\alpha}}\varphi}\,.
\end{equation}
Here we also neglected terms proportional to $\dot{\theta}$. We can further evaluate this expression analytically by employing the large $\varphi$ expansions of the potentials Eq.~\eqref{largephipot} and Eq.~\eqref{largephipothyp}. In this limit, we obtain
\begin{equation}\label{turnDEDE}
     \eta_{\perp}^{\rm mod}\approx \frac{72\sqrt{6\alpha}\ e^{-2\pi e^{-\sqrt{\frac{2}{3\alpha}}\varphi}}e^{\sqrt{\frac{2}{3\alpha}}\varphi}\ \sin(2\pi\theta) }{\sqrt{6\alpha}(\pi e^{\sqrt{\frac{2}{3\alpha}}\varphi}-3+6g_0)+12\varphi }\,,\quad \eta_{\perp}^{\rm hyper}\approx\frac{3\pi}{4}\beta^2 e^{-\frac{4 \varphi ^2}{3 \alpha }} \sin (2 \pi  \theta)\,,
\end{equation}
respectively. These are explicit functions of time once the approximate solutions $\theta\simeq {\rm const}$ and Eq.~\eqref{phiN} (or its analogous for the hyperbolic potential) are adopted. In Fig. \ref{turns}, we compare these estimates with the numerical results for the turning rate, further confirming that the assumption of effective single-field dynamics remains valid until the end of inflation, as the turning rate is extremely small throughout inflation in both models.

Although it is true that both these potentials are such that, in principle, they lead to the generic attractor behavior, this is subject to effects due to the multi-field dynamics, where, in particular, the dynamics along the $\theta$ direction may play a fundamental role. At large $\varphi$, the modular potential is double-exponentially suppressed in the $\theta$ direction (see Eq.~\eqref{largephipot}), so that the field $\theta$ is effectively massless. In turn, this leads to an initial ultra-slow-roll evolution for $\theta$, so that $\dot{\theta}\propto e^{-3Ht}$. After losing any initial velocity it might have, this field will therefore remain constant during inflation. In the hyperbolic manifold, this corresponds to geodesic motion; consequently, the background trajectory does not turn until the very end of inflation, as shown in Fig. \ref{dedekindbackgroundplots}. Therefore, although entropic perturbations are generated, these perturbations do not source the adiabatic ones, ensuring the universal large $N$ predictions Eqs.~\eqref{AsModpot}-\eqref{nsrmodpot} hold. The inflationary predictions of the model hence exactly match the ones from the cosmological attractors, as discussed earlier, up until the end of inflation.

However, towards the end of inflation and especially after this moment, $\varphi$ starts oscillating around its minimum, which then makes the $\theta$ direction to become apparent, so that eventually this field has to start rolling until it finally settles in one of its minima. Consequently, as the background trajectory needs to turn in order for $\theta$ to find its minimum, we expect that this turning may be responsible for sourcing the adiabatic perturbations from the entropic ones suddenly after the end of inflation.

\section{Inflationary evolutions of perturbations}\label{The adiabatic and entropic decomposition}

Having discussed the background evolution, we will now focus on the dynamics of linear perturbations by expanding the action up to the quadratic order in these quantities.

\subsection{Curvature and isocurvature perturbations}

Working in the spatially-flat gauge, where spatial scalar metric degrees of freedom are taken to be vanishing, the second-order Lagrangian for the perturbations $Q^I\equiv\phi^I-\bar{\phi}^I$ (where $\bar{\phi}^I$ denotes the background solution) is given by 
\begin{equation} \label{eomQ}
    \frac{{\cal L}(Q^{I})}{a^{3} }=\left[\left(G_{IJ}D_{t}Q^{I}D_{t}Q^{J}-\frac{(\partial Q^{I})\cdot(\partial Q^{J})}{a^{2}}\right)-M_{IJ}^{2}Q^{I}Q^{J}\right]\,.
\end{equation}
The components of the effective mass are defined as
\begin{equation} 
    M_{IJ}^{2}\equiv V_{;IJ}-\mathbf{R}_{IJKL}\dot{\phi}^{K}\dot{\phi}^{L}-\frac{1}{a^{3} }D_{t}\left(\frac{a^{3}}{H}\dot{\phi}_{I}\dot{\phi}_{J}\right)\,,
\end{equation}
where the covariant (second) derivative of the potential $V_{;IJ}$ and the Riemann tensor $\mathbf{R}_{IJKL}$ are constructed with the internal metric $G_{IJ}$ . The corresponding equations of motion, in Fourier space, read 
\begin{equation}
    D_{t}D_{t}Q^{I}+3HD_{t}Q^{I}+\left(\frac{k^{2}}{a^{2}}\delta^{I}_{
    J}+(M^2)^{I}_{J} \right)Q^{J}=0\,.
\end{equation}
To study the solutions of these equations, it is convenient to reparametrize the perturbations $Q^I$ into adiabatic and entropic components using the $n$-beins basis introduced in the previous section. Restricting to the two-field case, the adiabatic and entropic perturbations are then defined as
\begin{equation}
    Q_\sigma\equiv G_{IJ}e_\sigma^IQ^J\,, \quad Q_s\equiv G_{IJ}e_s^IQ^J\,,
\end{equation}
which correspond to perturbations respectively along and orthogonal to the background trajectory.

Within this decomposition, the comoving curvature perturbation $\cal R$, which is the relevant observable quantity, is then directly proportional to the adiabatic perturbation \cite{Bartolo_2001}, as
\begin{equation}
    {\cal R}\equiv H\frac{G_{IJ}\dot{\phi}^IQ^J}{G_{IJ}\dot{\phi}^I\dot{\phi}^J}=\frac{H}{\dot{\sigma}}Q_\sigma=\frac{1}{\sqrt{2\epsilon}}Q_\sigma\,.
\end{equation}
Analogously, we can define the isocurvature perturbation as
\begin{equation}
    {\cal S}\equiv\frac{H}{\dot{\sigma}}Q_s=\frac{1}{\sqrt{2\epsilon}}Q_s\,,
\end{equation}
so that we can study the evolution of scalar perturbations in terms of $\mathcal{R}$ and $\mathcal{S}$. Note that this is fully general for any two-field model of inflation, as we have not yet utilized the explicit form of the zweibeins.

Applying the adiabatic-entropic decomposition to Eq.~\eqref{eomQ}, we end up with the following equations of motion for $\mathcal{R}$ and $\mathcal{S}$
\begin{equation}\label{eomadi}
    \ddot{\cal R}+(3+\eta_{H})H\dot{\cal R}+\frac{k^2}{a^2}{\cal R}=\frac{1}{a^3\epsilon}\frac{d}{dt}\left(2\epsilon a^3H\eta_\perp{\cal S}\right)\,,
\end{equation}
\begin{equation}\label{eomiso}
    \ddot{\cal S}+(3+\eta_{H})H\dot{\cal S}+\left(\frac{k^2}{a^2}+m_\mathcal{S}^2\right){\cal S}=-2H\eta_\perp\dot{{\cal R}}\,,
\end{equation}
where we introduced the mass of the isocurvature perturbation
\begin{align}
    m_\mathcal{S}^2&\equiv V_{;ss}-V_{;\sigma\sigma}+\epsilon H^{2}\mathbf{R}+\frac{1}{a^3}\frac{d}{dt}\left(\frac{a^3\dot{\sigma}^2}{H}\right)\,,
\end{align}
where $\mathbf{R}$ is the Ricci scalar of the field space metric, and the covariant derivative of the potential in the adiabatic-entropic basis is given by
\begin{equation}
    V_{;ij}\equiv e_i^Ie_j^JV_{;IJ}\,.
\end{equation}
Notice that the curvature perturbation $\cal R$ is exactly massless, so that it would go to a constant on super-horizon scales ($k\ll aH$) in the absence of a source term on the right side of its equation. In turn, the isocurvature perturbation $\cal S$ is, in general, massive and thus will either be suppressed or enhanced (on super-horizon scales) depending on the sign of its mass. Moreover, from Eqs.~\eqref{eomadi}-\eqref{eomiso} it is apparent that there exists an intrinsic relationship between these two types of perturbations. As they source each other, their evolution is coupled. However, notice that both sources depend not only on $\cal R$ or $\cal S$ but also on the turning rate. Therefore, in order to have an efficient sourcing of curvature perturbation, the background trajectory must experience a non-negligible turning rate.

\subsection{Super-horizon evolution}

\noindent
On super-horizon scales ($k\ll aH$), the dynamics of $\cal R$ and $\cal S$ can be further simplified. In this limit, Eq.~\eqref{eomadi} can be written as a total derivative \cite{Pinol:2021nha,Iacconi:2021ltm}
\begin{equation}
    \frac{1}{a^3\epsilon}\frac{d}{dt}\left(a^3\epsilon\dot{\cal R}-2\epsilon a^3H\eta_\perp{\cal S}\right)={\cal O}\left(\frac{k^2}{a^2}\right),
\end{equation}
which can be integrated directly as
\begin{equation} \label{Rdot}
    \dot{{\cal R}}=2H\eta_\perp{\cal S}+\frac{C(k)}{a^3\epsilon}\,,
\end{equation}
where the term proportional to the $k$-dependent integration constant $C(k)$ becomes quickly negligible as it is suppressed by $a^3$. Substituting this result back in Eq.~\eqref{eomiso} leads to
\begin{equation}\label{SuperHorizonS}
    \ddot{\cal S}+(3+\eta_{H})H\dot{\cal S}+m_{\mathcal{S},{\rm eff}}^2{\cal S}={\cal O}\left(\frac{k^2}{a^2}\right)\,
\end{equation}
where we defined the effective isocurvature mass \begin{equation}\label{effectiveisomass}
    m_{\mathcal{S},{\rm eff}}^2\equiv m_\mathcal{S}^2+4H^2\eta_\perp^2\,.
\end{equation} 
We have thus obtained that, on super-horizon scales, the isocurvature perturbation $\mathcal{S}$ decouples from the curvature one and evolves independently\footnote{Ideally, one can solve Eqs.~\eqref{Rdot}-\eqref{SuperHorizonS} by setting the initial condition for $\cal R$ and $\cal S$ at horizon crossing ($k=aH$). In practice, the term proportional to $k^2/a^2$ is not yet negligible at this moment, so that the initial conditions should be set instead a few $e$-folds later.}. On the other hand, it directly sources $\mathcal{R}$ via Eq.~\eqref{Rdot} in the presence of a non-zero turning rate $\eta_\perp$, which is precisely the phenomenon we want to investigate in this work.

\begin{figure}[t]
       \begin{subfigure}[b]{0.45\textwidth}
        \centering
        \includegraphics[width=1\linewidth]{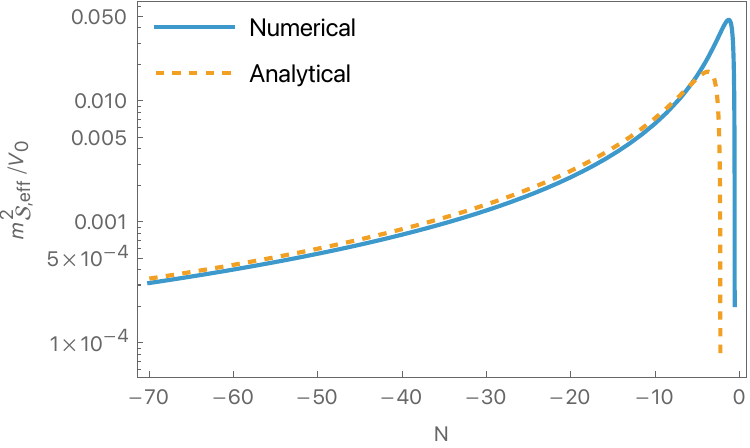}  
        \label{fig:plot2b}
    \end{subfigure}
    \hfill
    \begin{subfigure}[b]{0.45\textwidth}
        \centering
        \includegraphics[width=1\linewidth]{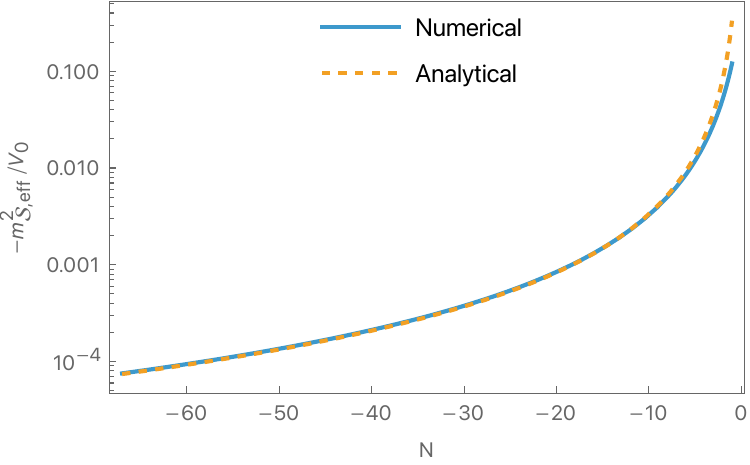} 
        \label{fig:plot4b}
    \end{subfigure}
    \caption{\footnotesize {Plot of the effective isocurvature mass squared $m_{\cal{S}, \rm eff}^{2}$ (in units of $V_0$) for the modular (left panel) and hyperbolic (right panel) models. In both plots, we compare the analytical formulas Eqs.~\eqref{massdedekind} and \eqref{masstanh} in blue together with the numerical evaluation of Eq.~\eqref{effectiveisomass}, for each model, in yellow. The main result from these plots is the difference in sign of the isocurvature effective mass for both models, suggesting a different behavior of this perturbation.}}
    \label{MassPlot}
\end{figure}

We can gain a clearer understanding of the behavior of the isocurvature perturbations by looking at their mass. Under the slow-roll approximation, the  effective isocurvature mass reads 
\begin{equation}\label{massformula}
    m^2_{{\cal S}, \rm eff}\approx  \left(1-\frac{2}{9\alpha}\right)\frac{(V_\varphi)^2}{3H^2}-\sqrt{\frac{2}{3\alpha}} V_\varphi-V_{\varphi\varphi}\,,
\end{equation}
where we used the following relation during inflation 
\begin{equation}
    \dot{\sigma}^2\simeq\dot{\varphi}^2\simeq\frac{(V_\varphi)^2}{9H^2}\,.
\end{equation}
We can further develop Eq.~\eqref{massformula} for each model under consideration by employing the large $\varphi$ (large $N$) expansions in Eqs.~\eqref{largephipot} and \eqref{largephipothyp}, obtaining 
\begin{equation}\label{massdedekind}
   m^{2}_{\cal{S},\rm eff}\approx -\frac{3\alpha V_{0}}{2g_0 N^{2}}\log\left[\frac{4g_{0}N}{\pi\alpha}  \right] -\frac{V_{0}}{3N^{2}}+\frac{9\alpha V_{0}}{4 g_0 N^{2}} \,, \quad \mbox{(modular potential)}
\end{equation}
and
\begin{equation}\label{masstanh}
    m^{2}_{\cal{S},\rm eff}\approx -\frac{V_{0}}{3N^{2}}\,, \quad \mbox{(hyperbolic potential)}
\end{equation}
respectively. These share the important feature that during the large part of inflation, the effective entropic mass is suppressed as $1/N^2$ -- with all $1/N$ terms from Eq.~\eqref{massformula} canceling out. This component is therefore also effectively massless during the largest part of inflation, and only grows to significant values towards its end. This is illustrated in Fig.~\ref{MassPlot}, which displays both these analytical formulas and the numerical computation of the effective isocurvature mass. As we will see, the signs of the remaining contributions to the effective mass (which are positive for the modular and negative for the hyperbolic potential) are directly related to the behaviour of the isocurvature mode on super-horizon scales.

Interestingly, no assumption on $\epsilon$ is made in the previous derivation of the superhorizon evolution in Eqs.~\eqref{Rdot} and \eqref{SuperHorizonS}, so that these are still valid for $\epsilon>1$. Once the initial conditions are properly set on super-horizon scales, this set of equations provides a better understanding of the behavior of curvature and isocurvature perturbations, even beyond the end of inflation.

\subsection{Quantization and power spectra}\label{Canonical variables, quantization and power spectra}

\noindent In order to solve the equations of motion \eqref{eomadi}-\eqref{eomiso}, initial conditions for the fields have to be set in the infinite past, when all the relevant modes are deep on sub-horizon scales with $k\gg aH$. The proper way of doing so is by accounting for the quantum nature of the perturbations, setting their non-vanishing initial amplitudes as dictated by the laws of quantum field theory. The first step is to canonically normalize the fields in the following way
\begin{equation}\label{canonicalredefinitiontensors}
    \Delta\equiv\begin{pmatrix}
        {\cal R}\\
        {\cal S}\\
    \end{pmatrix}={\cal M}\cdot\begin{pmatrix}
        {\cal R}_c\\
        {\cal S}_c
    \end{pmatrix}
    \equiv{\cal M}\cdot\Delta_c\,,
\end{equation}
where we label the doublets of physical and canonically-normalized fields as $\Delta$ and $\Delta_c$, respectively, which are related by the rotation matrix $\mathcal{M}$. By construction, this is diagonal and proportional to the identity as
\begin{equation} \label{rotationM}
   {\cal M}=\frac{H}{a\dot{\sigma}}\mathbb{I}\,.
\end{equation}
The quantization of the fields is then done by promoting $\Delta_c$ to a quantum operator
\begin{equation}
    \Delta_{c,i}\quad\rightarrow\quad\hat{\Delta}_{c,i}={\cal D}_{ij}\hat{a}_{j}+{\cal D}^{*}_{ij}\hat{a}^{\dagger}_{j}\,,
\end{equation}
which is a linear combination of two independent sets of creation/annihilation operators satisfying the canonical commutation relations
\begin{equation}
    \left[\hat{a}_i(\vec{k}),\hat{a}_j^\dagger(\vec{k}')]\right]=\delta(\vec{k}-\vec{k}')\delta_{ij}\,.
\end{equation}
The $n^2=4$ mode-functions (with $n$ number of fields) ${\cal D}_{ij}$ satisfy the classical equations of motion. The system of equations is then solved by setting the Bunch-Davies initial conditions in the infinite past 
\begin{equation}\label{initialtensors}
   \sqrt{2k} {\cal D}_{\rm in}=\begin{pmatrix}
        e^{-ik\eta} & \quad 0\\
        0 & \quad e^{-ik\eta}
    \end{pmatrix}_{\eta=\eta_{\rm in}},  \qquad \sqrt{2k}(\partial_{\eta}{\cal D})_{\rm in}=-ik\begin{pmatrix}
        e^{-ik\eta} & \quad 0\\
        0 & \quad e^{-ik\eta}
    \end{pmatrix}_{\eta=\eta_{\rm in}}\,,
\end{equation}
where $\eta$, not to be confused with the Dedekind function $\eta_{D}$, is the conformal time and the initial time $\eta_{\rm in}$ is such that $-k\eta_{\rm in}\rightarrow\infty$ for each relevant mode. Once the solutions for the mode-functions ${\cal D}$ are obtained, it is possible to compute correlators of $\hat{\Delta}_{i}$ as
\begin{equation}
    \langle\hat{\Delta}_i({\vec{k}})\hat{\Delta}_j({\vec{k}'})\rangle={\cal M}_{ik}{\cal M}_{j\ell}\langle\hat{\Delta}_{c,k}({\vec{k}})\hat{\Delta}_{c,\ell}({\vec{k}'})\rangle=\delta(\vec{k}-\vec{k}'){\rm Re}\left[{\cal M}_{ik}{\cal M}_{j\ell}{\cal D}_{km}{\cal D}^*_{\ell m}\right]\, .
\end{equation}
In the last step, we used the canonical commutation relations. The dimensionless power spectrum is then defined as
\begin{equation}
    \langle\hat{\Delta}_i({\vec{k}})\hat{\Delta}_j({\vec{k}'})\rangle\equiv\frac{2\pi^2}{k^3}\delta(\vec{k}-\vec{k}'){\cal P}_{ij}(k)\, .
\end{equation}
Combining the previous results, we obtain
\begin{equation}
    {\cal P}_{ij}(k)=\frac{k^3}{2\pi^2}{\cal M}_{ik}{\cal M}_{j\ell}{\cal D}_{km}{\cal D}^*_{\ell m}\,,
\end{equation}
where in particular we have the self-correlations ${\cal P}_{\cal R}\equiv{\cal P}_{11}$ and ${\cal P}_{\cal S}\equiv{\cal P}_{22}$, while the cross-correlation between the two fields is given by the off-diagonal component ${\cal P}_{\times}\equiv{\cal P}_{12}={\cal P}_{21}$. Inserting the explicit expression of the matrix $\mathcal{M}$ in Eq.~\eqref{rotationM} we finally find the power spectra
\begin{equation}
    {\cal P}_{\cal R}(k)=\frac{k^3}{2\pi^2}\frac{H^2}{a^2\dot{\sigma}^2}{\cal D}_{1m}{\cal D}^*_{1m}\,, \quad {\cal P}_{\cal S}(k)=\frac{k^3}{2\pi^2}\frac{H^2}{a^2\dot{\sigma}^2}{\cal D}_{2m}{\cal D}^*_{2 m}\,,
\end{equation}
as well as the cross-correlation spectrum
\begin{equation}
    {\cal P}_{\times}(k)=\frac{k^3}{2\pi^2}\frac{H^2}{a^2\dot{\sigma}^2}{\cal D}_{1m}{\cal D}^*_{2m}\,.
\end{equation}
In what follows, we will focus on the information on the two power spectra.

\subsection{Sourcing of curvature from isocurvature perturbations}

As we have seen in the previous sections, in multi-field models of inflation, adiabatic and entropic perturbations source each other. This suggests a natural decomposition of the curvature and isocurvature power spectra as
\begin{equation}\label{SplittingPS}
    \mathcal{P}_{\mathcal{R}}(k)=\mathcal{P}_{\mathcal{R}}^{0}+\mathcal{P}_{\mathcal{R}}^{s}\,, \quad
    \mathcal{P}_{\mathcal{S}}(k)=\mathcal{P}_{\mathcal{S}}^{0}+\mathcal{P}_{\mathcal{S}}^{s}\,,
\end{equation}
where we identify the vacuum and sourced components with $0$ and $s$, respectively. As discussed, the sourcing depends linearly on the turning rate, which we found to be negligible during inflation, as visible in Fig. \ref{turns}. Because of this, we expect the dynamics of curvature and isocurvature perturbations to remain decoupled during inflation and in particular, on super-horizon scales, to be solely determined by the effective mass of these quantities (which for $\mathcal{R}$ is vanishing and for $\mathcal{S}$ behaves as $1/N^2$).

\begin{figure}[t!]
    \begin{subfigure}[b]{0.45\textwidth}
        \centering
        \includegraphics[width=1\linewidth]{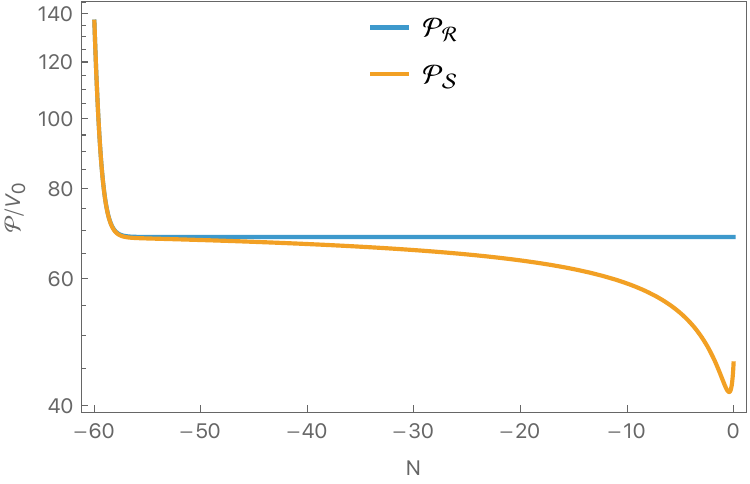}  
    \end{subfigure}
    \hfill
    \begin{subfigure}[b]{0.45\textwidth}
        \centering
        \includegraphics[width=1\linewidth]{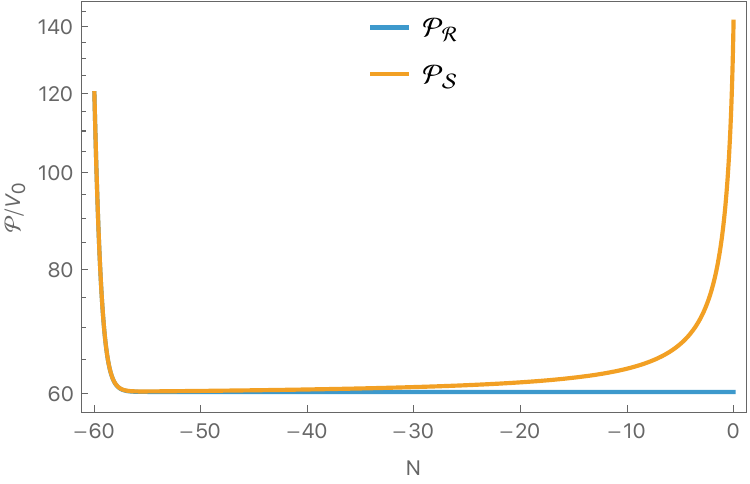} 
    \end{subfigure}
    \caption{\footnotesize {Plot of the curvature and isocurvature power spectra (in units of $V_0$) for the modular (left panel) and the hyperbolic (right panel) potentials. In both cases, the curvature power spectrum follows its own evolution on superhorizon scales and is thus mainly dominated by its vacuum component, which remains constant after horizon crossing, while the isocurvature one feels the effect of its effective mass squared (see Eqs.~\eqref{massdedekind}-\eqref{masstanh} and Fig. \ref{MassPlot}). Note that the different sign of the latter in both cases leads to a small enhancement or suppression, thus explaining the small differences between the two power spectra as inflation proceeds.}}
    \label{inflaadiabatic}
\end{figure}

The numerical solutions for the power spectra of curvature and isocurvature perturbations, for both choices of the potential, are presented in Fig. \ref{inflaadiabatic} where we show the evolution of a single CMB mode, crossing the horizon at $N_{\rm hc}=60$. 
We confirm that in both cases, the curvature mode exhibits no significant enhancement from the isocurvature one, as expected due to the strong suppression of the turning rates. In contrast, one key difference lies in the behavior of the isocurvature perturbation on super-horizon scales, which is a direct consequence of the opposite sign of the effective mass squared (see Fig. \ref{MassPlot}).

In conclusion, during inflation, both models exhibit similar qualitative behaviors. As the field trajectory remains geodesic, the $\theta$ direction stays suppressed, and thus the turning during inflation is negligible. Therefore, the sourcing of isocurvature into curvature perturbations is not efficient, resulting in the total curvature spectrum coinciding with the one obtained in the effective single-field approximation. This is in contrast to other two-field models in the literature, where the turning takes place already during inflation, therefore leading to an active sourcing of curvature perturbation from isocurvature ones \cite{Fumagalli:2019noh,Christodoulidis:2019jsx,Bjorkmo:2019fls,Achucarro:2019mea,Achucarro:2019pux,Iarygina:2020dwe,Iacconi:2021ltm,Iarygina:2023msy}.


\section{Post-inflationary evolution of perturbations}\label{Postinflationary evolution}

\subsection{Enhancement of adiabatic perturbations}

In the previous section, we confirmed numerically that, for the models under consideration, the turning rate is suppressed during inflation, and thus the sourcing of curvature from isocurvature perturbations is negligible. Notice, however, that no slow-roll assumptions have been made in the derivation of equations for the perturbations Eqs.~\eqref{eomadi}-\eqref{eomiso} and hence we can, in principle, extend them beyond the end of inflation. Likewise, the super-horizon limit in Eqs.~\eqref{Rdot}-\eqref{SuperHorizonS} remains a valid approximation as we are interested in the evolution of modes that crossed the horizon long before the end of inflation.

The motivation of going beyond $\epsilon=1$ is understood by looking at the background evolution in Figures \ref{dedekindbackgroundplots} and \ref{tanhbackground}, where, for both models under consideration, the turning in field space always happens after the end of inflation. Indeed, the turn rate becomes ${\mathcal O} (10)$ approximately one $e$-fold after the end of inflation, precisely at the moment the fields start to oscillate, and later grows until reaching ${\mathcal O}  (100)$ at $N\sim 3$. This suggests that at this stage, the isocurvature perturbation $\cal S$ can be efficiently converted into curvature $\cal R$ due to the turning rate being no longer negligible (see the superhorizon limit Eq.~\eqref{Rdot}). This, along with the fact that $\cal {S} $ remains at a comparable amplitude with respect to $\cal{R}$ (see Fig.  \ref{inflaadiabatic}), leads to a net sourcing of $\mathcal {R} $ after the end of inflation.

\begin{figure}[t]
    \centering
       \begin{subfigure}{0.45\textwidth}
        \centering
        \includegraphics[width=1\linewidth]{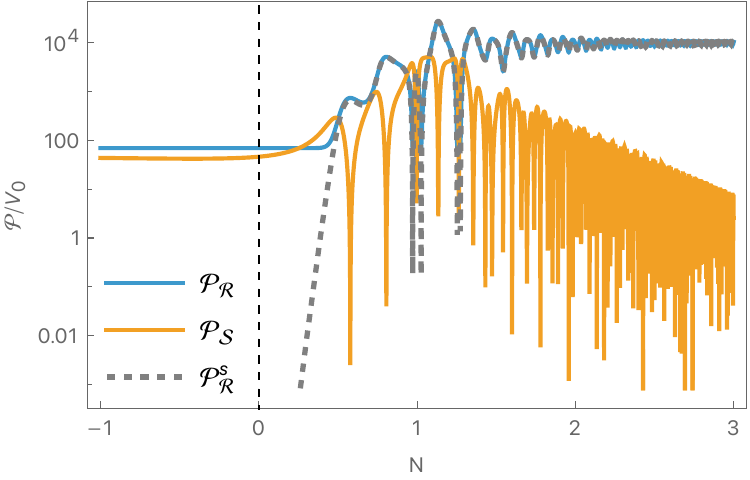}  
        \end{subfigure}
    \hfill
    \begin{subfigure}{0.45\textwidth}
        \centering
        \includegraphics[width=1\linewidth]{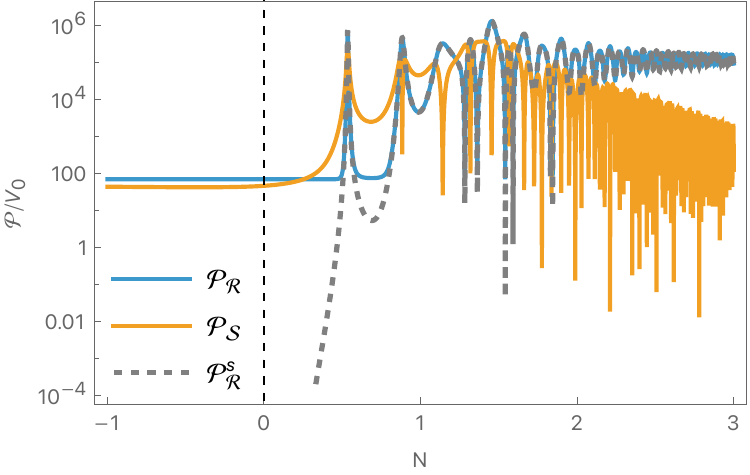} 
        \end{subfigure}
    \vfill
    \begin{subfigure}{0.45\textwidth}
        \centering
        \includegraphics[width=1\linewidth]{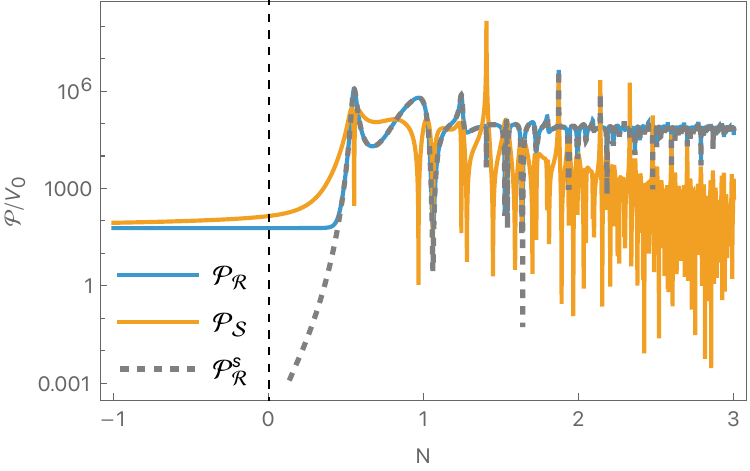} 
        \end{subfigure}
    \caption{\footnotesize {Plot of the post-inflationary curvature and isocurvature power spectra (in units of $V_0$) for the modular invariant Eq.~\eqref{dedekindmodel} (upper panels) and for the hyperbolic model Eq.~\eqref{tanhmodel} (bottom panel). For both models, we set $3\alpha=1$, while the initial condition is given by $\theta_{\rm in}=0.3$ (top left) and $\theta_{\rm in}=0.01$ (top right and bottom). The dashed vertical line indicates the end of inflation. In these plots, we clearly see how the sourcing of curvature from isocurvature perturbations becomes non-negligible beyond the end of inflation. More interestingly, we find that, soon after the end of inflation, the sourcing from isocurvature into curvature perturbations comes to a halt, reaching a new amplitude for the curvature power spectrum. These plots also include the evolution of the sourced component of the curvature power spectrum $\mathcal{P}_{\mathcal{R}}^{s}$ (dashed gray), which is negligible during inflation but later becomes the dominant contribution after inflation ends. }}
\label{dedekindplateau1}
\end{figure}

Our strategy then boils down to evolve numerically the system of scalar perturbations after the end of inflation until $N\simeq3$ (we discuss the implications and validity of this analysis in Sub-Section \ref{Reheating considerations}), which allows us to capture the sourcing of $\cal R$ due to its coupling with $\cal S$. We report the numerical results of this in Fig. \ref{dedekindplateau1}. Notice that, even though we are showing only the solutions around and after the end of inflation, these should be interpreted as extensions of the plots reported in the previous section (see Fig. \ref{inflaadiabatic}). In particular, we emphasize that the initial conditions for the modes have been properly set in the deep sub-horizon limit and evolved using the full system of equations, as detailed in subsection \ref{Canonical variables, quantization and power spectra}.

As expected, after the end of inflation, we find a net enhancement of $\cal R$ as sourced by $\cal S$ due to the large turning in field space. Interestingly, the sourced component of $\cal R$, which is the dominant one after the sourcing, approaches a constant value. Furthermore, from Fig. \ref{dedekindplateau1} we also notice that the curvature power spectrum reaches its new enhanced value in the span of $\mathcal{O}(1)$ $e$-folds after the end of inflation, justifying stopping the numerical evaluation at $N\simeq3$.

The appearance of the new plateau can be understood in the following way. After reaching its maximum, the source term $\eta_{\perp}\cal{S}$ in Eq.~\eqref{Rdot} starts to oscillate periodically, which effectively stops the sourcing of curvature perturbation (given by the integral of the source term). Moreover, the isocurvature power is dropping during these oscillations, and hence the amplitude of the sourcing is exponentially decaying.

\begin{figure}[t]
    \centering
       \begin{subfigure}{0.45\textwidth}
        \centering
        \includegraphics[width=1\linewidth]{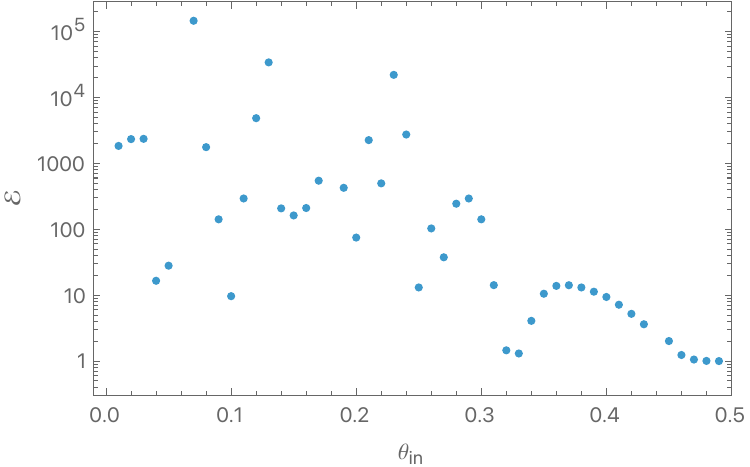}  
        \end{subfigure}
    \hfill
    \begin{subfigure}{0.45\textwidth}
        \centering
        \includegraphics[width=1\linewidth]{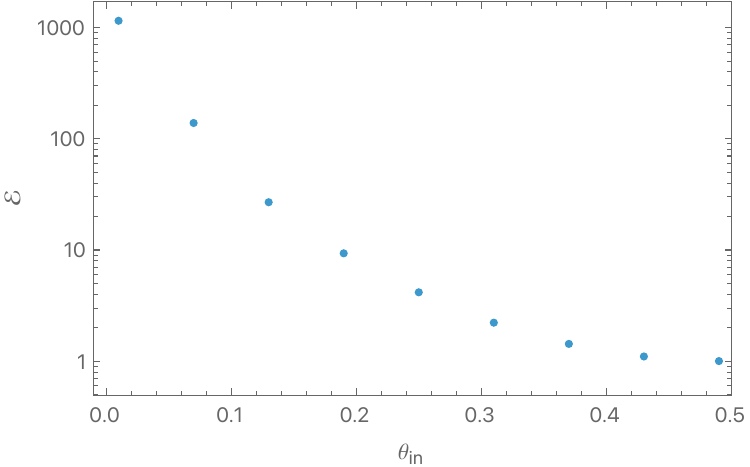} 
        \end{subfigure}\caption{\footnotesize {Plot of the post-inflationary enhancement Eq.~\eqref{defenhancement} of the curvature power spectrum for several initial values $\theta_{\rm in}$. In the left panel, we show the dependence of the power spectrum enhancement in the case of the modular-invariant model Eq.~\eqref{dedekindmodel}. In the right panel, we show this dependence for the hyperbolic one Eq.~\eqref{tanhmodel}. These plots show that the enhancement of the curvature power is very sensitive to even small changes of the initial values of $\theta$ in the case of the modular potential \eqref{dedekindmodel}, suggesting a possibility of large non-Gaussianity  for $\theta_{\rm in}  \lesssim 0.4$. In contrast, in the hyperbolic model \eqref{tanhmodel}, we find a smooth, monotonously decreasing enhancement for increasing $\theta_{\rm in}$. In both cases, the point $\theta_{\rm in}=0.5$ is special as this is the value for the minima of the $\theta$ field. Therefore, if the field starts with this particular value, no turning in field space happens, and thus no sourcing of curvature power is present. }}
    \label{enhancement}
\end{figure}

We shall now discuss the effects of this post-inflationary enhancement. We begin by defining the enhancement of the scalar power spectrum as 
\begin{equation}\label{defenhancement}
    {\cal E}\equiv\frac{{\cal P}^{\rm s}_{{\cal R}}}{{\cal P}^{0}_{{\cal R}}}\,,
\end{equation}
which we evaluate after inflation at $N\simeq3$, when the plateau is fully developed. We would like to emphasize that this quantity is heavily dependent on the initial condition $\theta_{\rm in}$ and must be calculated on a case-by-case basis; this is illustrated in Fig. \ref{enhancement} for both models. However, a first naive expectation is as follows.

Towards the end of inflation, after being frozen at its initial value $\theta_{\rm in}$, the field $\theta$ starts to look for its minimum, which, for both models under consideration, is located at $\theta=0.5$. Therefore, the farther away $\theta$ starts from its minimum, the longer it will take to reach it. The background trajectory will then experience an increased turning compared to the scenario when $\theta$ starts very close to its minimum. Consequently, we expect the enhancement to be decreasing with the distance between $\theta_{\rm in}$ and $0.5$, and in particular to be vanishing in the limiting case where $\theta_{\rm in}=0.5$. This is confirmed in the case of the hyperbolic model (see the right panel of Fig.  \ref{enhancement}) where we plot the dependence of the enhancement in Eq.~\eqref{defenhancement} with respect to $\theta_{\rm in}$.

In the case of the modular model, we encounter a somewhat different situation. The enhancement of curvature perturbations follows the naive expectation described above only for $\theta_{\rm in} \gtrsim 0.4$. We attribute the difference between the results in the modular and hyperbolic models to the different and richer structure in the $SL(2,\mathbb Z)$ potential and on its minima. This strong dependence on initial conditions may be pointing towards a large non-Gaussian signal in CMB observables \cite{Peterson:2010mv,Iarygina:2023msy}  for $\theta_{\rm in} \lesssim 0.4$  in the case of the modular-invariant model Eq.~\eqref{dedekindmodel}. We leave an investigation of a possible non-Gaussianity for future work.\footnote{Note that one can suppress non-Gaussianity (and isocurvature perturbations) in the models with axion stabilization, see \cite{Carrasco:2025rud} and  Appendix \ref{A comment on the stabilization of the modular models}.}


\subsection{Inflationary observables}

Most inflationary observables follow from the curvature spectrum that sources structure formation, and hence have to be evaluated taking into account the enhancement that occurs after the end of inflation. This concerns firstly the amplitude of this power spectrum, with  
\begin{equation}
    A^{\rm post}_{s} = \mathcal{E} A^{\rm end}_{s} \,,   
\end{equation}
relating the amplitude at the end of inflation to the amplitude after the post-inflationary enhancement (where the precise coefficient of this proportionality relation depends on $\theta_{\rm in}$ and must be calculated case by case). Importantly, $A_{s}^{\rm post}$ has to agree with the COBE normalization by correspondingly fixing the potential scale $V_{0}$.

\begin{figure}[t]
    \centering
       \begin{subfigure}{0.45\textwidth}
        \centering
        \includegraphics[scale=0.5]{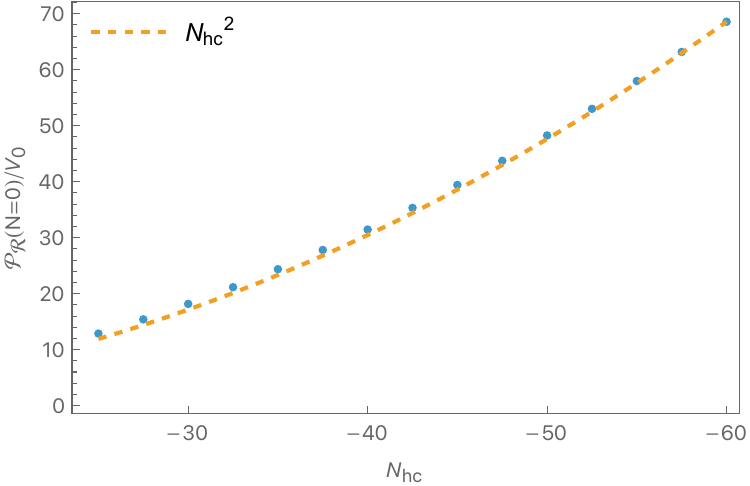}  
        \end{subfigure}
    \hfill
    \begin{subfigure}{0.45\textwidth}
        \centering
        \includegraphics[scale=0.54]{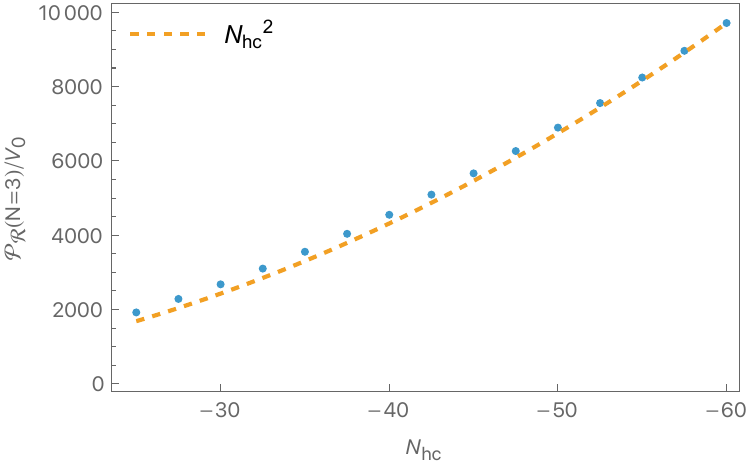} 
        \end{subfigure}
    \vfill
    \begin{subfigure}{0.45\textwidth}
        \centering
        \includegraphics[scale=0.5]{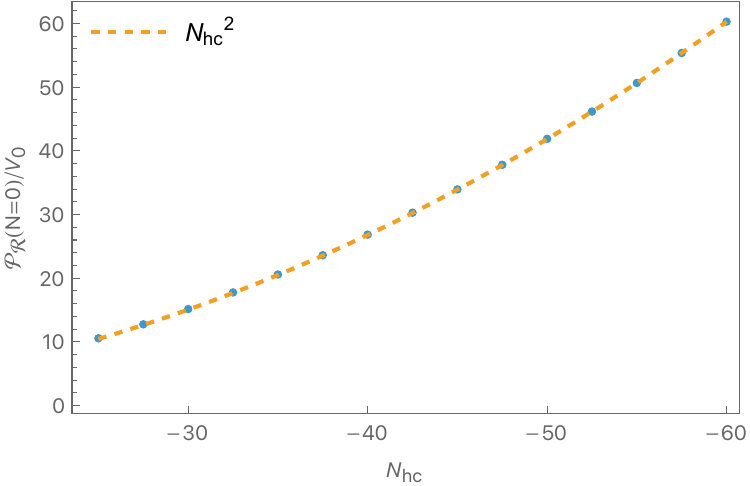}  
        \end{subfigure}
    \hfill
    \begin{subfigure}{0.45\textwidth}
        \centering
        \includegraphics[scale=0.54]{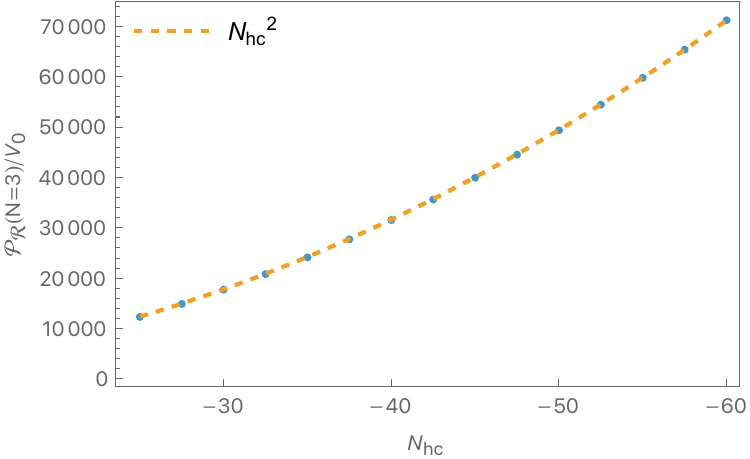} 
        \end{subfigure}
    \caption{\footnotesize {Plot of the scaling of the power spectrum of curvature perturbations (in arbitrary units) for the model in Eq.~\eqref{dedekindmodel} (upper panels) and in Eq.~\eqref{tanhmodel} (lower panels) as a function of the time of horizon crossing, identified with $N_{\rm hc}$. In the left panels we show the scaling of the amplitude of the power spectra evaluated at the end of inflation $(N=0)$, which is in agreement with the $\alpha$-attractor expression $A_{s}\sim N_{\rm hc}^{2}$. On the right panel instead, we show this same scaling but evaluating the amplitude of the power spectra three $e$-folds after the end of inflation $(N=3)$. This figure shows that, in both models, the $N_{hc}^{2}$ scaling of the amplitude of the power spectrum remains a good description even after the end of inflation.}}
    \label{scaling}
\end{figure}

Secondly, the spectral index observed in the CMB is set by the scale dependence of $A^{\rm post}_{s}$. We show in Fig. \ref{scaling} that, for both models, the amplitude of the curvature perturbation after the enhancement $A^{\rm post}_{s}$ still follows the same $N_{\rm hc}^2$-scaling as the amplitude at the end of inflation. This important result of our investigation is a direct consequence of the hyperbolic geometry, the structure of the $\alpha$-attractor potentials of the field $ \varphi$, and the extreme flatness of the axion potential during inflation  \cite{Kallosh:2024whb}.
 
To understand this result qualitatively, we note that the kinetic term of the field $\theta$ in Eq.~\eqref{ModularKineticterm}  differs from the familiar kinetic term of a canonically normalized field $\frac12 \partial_\mu\theta\partial^\mu\theta$  by the exponential coefficient $\frac{3\alpha}{2} \exp\left({-2\sqrt{\frac{2}{3\alpha}}\vp}\right)$. In the theory of $\alpha$-attractors, this coefficient is directly related to the number of $e$-folds of inflation at $\varphi \gg \sqrt{\frac{3\alpha}{2}}$:  
\be
  \varphi \simeq \sqrt{3 \alpha \over 2} \log(|N|) \,,
\ee
up to a small and constant, model-dependent shift of $\varphi$ (see e.g.~\eqref{phiN})\footnote{{We remind the reader that we have adopted the convention $N<0$ during inflation}.}.
This means that the kinetic term of the field $\theta$ can be represented as follows:
 \begin{equation}
  {1\over  N^{{2}}} \left({3\alpha\over 2}\right) \frac12 \partial_\mu\theta\partial^\mu\theta \,.
  \label{Ndelta}
\end{equation}
Consider now the early stages of inflation where $|N| \gg 1$. Then during a typical Hubble time $H^{{-1}}$ the value of $|N|$ remains almost constant: $\Delta N = 1 \ll |N|$. Therefore, one may argue that in this limit, one may simply treat the coefficient in Eq.~\eqref{Ndelta} as a constant and use it to rescale the variable $\theta$ to make it canonically normalized:
\be\label{normal}
\tilde\theta = {1\over  |N|}  \left({3\alpha\over 2}\right)^{1/2} \theta \,.
\ee
The amplitude of quantum fluctuations of the field $\tilde\theta$ generated during the Hubble time should then be given by the familiar expression $H/(2\pi)$. If this is the case, then Eq.~\eqref{normal}  implies that
\be\label{ampth}
\delta\theta = {H\over 2\pi} \left(\frac{2}{3\alpha}\right)^{1/2}N_{\rm hc} \,,
\ee
where $N_{\rm hc}$ marks the horizon-crossing time of the perturbation mode $\delta\theta$. Therefore, as explained in Ref.~\cite{Kallosh:2024whb}, the amplitude of isocurvature perturbations both during and at the end of inflation is proportional to  $(\delta \theta)^2 \propto N_{\rm hc}^{2}$. This large factor is the primary reason why significant post-inflationary amplification of perturbations can occur in our scenario.

At the end of inflation, both the adiabatic and isocurvature modes therefore have the same $N_{\rm hc}^{2}$ scaling. Regardless of the details of the post-inflationary evolution, which we studied in this paper, the resulting amplitude $A^{\rm post}_{s}$ will therefore have the same scaling -- it may have a complicated dependence on the initial value of the field $\theta$, but the overall factor $N_{\rm hc}^{2}$ is universal. From this result, one can immediately conclude that the spectral index remains the same as in the single field $\alpha$-attractor models,  
\begin{equation}
    n_{s}^{\rm post}=n_{s}^{\rm end} \approx 1-\frac{2}{N_{\rm hc}} \,. \label{spectral index}
\end{equation}
One should note, however, that all general results in the theory of $\alpha$-attractors  are only valid in the leading approximation in the $1/N_{\rm hc}$ expansion. This may explain a slight deviation of the results of our numerical investigation shown in the upper panel of Fig. \ref{scaling} from the exact $N_{\rm hc}^{2}$ scaling and, similarly, may lead the two spectral indices in Eq.~\eqref{spectral index} to differ by small terms $\mathcal{O}(1/N_{\rm hc}^{2})$.

The enhancement of the scalar power spectrum then translates into a suppression of the tensor-to-scalar ratio 
\begin{equation}\label{decreaser}
    r^{\rm post}=\frac{r^{\rm end}}{\mathcal{E}} \approx \frac{12 \alpha}{\mathcal{E} N_{\rm hc}^2}\,,
\end{equation}
as the tensor power spectrum remains unaffected. The post-inflationary power enhancement thus leads to a suppression of the tensor-to-scalar ratio compared to the standard $\alpha$-attractor prediction.

\subsection{Reheating considerations}\label{Reheating considerations}
Throughout this work, we have neglected effects related to post-inflationary reheating. After inflation, the inflaton field $\vp$ oscillates with a frequency equal to its mass $m$. The basic, perturbative mechanism of reheating involves an oscillating scalar field being represented as a collection of particles that may decay into lighter ones. For example, a scalar field $\vp$ may interact with light fermions with a coupling $g \vp \bar \psi \psi$. This leads to decay of the scalar field to fermions with the decay rate $\Gamma = {g^{2} \over 4\pi} m$ \cite{Linde:1990flp}.

 We can effectively take into account this decay by modifying the background equations of motion Eq.~\eqref{phiEOM} and Eq.~\eqref{ThetaEOM} to include a dissipation term $\Gamma$:
 \begin{equation}\label{phiEOMGAMMA}
\begin{split}
    \ddot{\varphi}+(3H+\Gamma)\dot{\varphi}+\frac{\partial V(\varphi,\theta)}{\partial \varphi}+\sqrt{\frac{3\alpha}{2}}e^{-2 \sqrt{\frac{2}{3\alpha}}\varphi} \dot{\theta }^2=0\,, \\
    \ddot{\theta}+(3H+\Gamma)\dot{\theta}-\sqrt{\frac{2}{3\alpha}}\dot{\theta}\dot{\varphi}+\frac{2e^{2 \sqrt{\frac{2}{3\alpha}} \varphi}} {{3 \alpha }} \frac{\partial V(\varphi,\theta)}{\partial \theta}=0\,.
\end{split}
\end{equation}
The most important part of the processes studied in this paper occurs during the first few oscillations, with a frequency of $\mathcal{O}(m)$.  At the end of inflation, $H = \mathcal{O}(m)$. Therefore $\Gamma \sim  {g^{2} \over 4\pi} H$.

This implies that if the fields $\vp$ and $\theta$ are weakly coupled to other fields, $g\ll 1$, their decay rate is small, and one can safely ignore $\Gamma$ as compared to $H$ in these equations during the first few post-inflationary oscillations.

To make this analysis more quantitative, we will make an order-of-magnitude estimate of the properties a given perturbative reheating process must fulfill in order for us to safely neglect it at $N\sim 3$. In order to do this, we first normalize the enhanced amplitude of the curvature power to its measured value by fixing $V_{0}$. Note that this is initial-condition dependent because of the same dependence on the enhancement of curvature perturbations. We then numerically calculate $H_{p}\equiv H(N=3)$ with which we set an upper bound on the decay rate $\Gamma <H_{p}/10 $ up until which we assume we can safely neglect it. Finally, we relate the interaction rate to the reheating temperature as  \cite{McDonald:1999hd,Kolb:2003ke}
\begin{equation}
    T_{\rm re}\approx\sqrt{\Gamma}\,.
\end{equation} 
This process leads to the following estimates for the initial conditions considered in the previous plots:
\begin{equation}\label{Tre}
\begin{aligned}
    &\mbox{Modular model:} && \theta_{\rm in} = 0.3\,,         && V_0 \simeq 10^{-13}\,,            && H_p \simeq 10^{-9}\,,
    && \Gamma \simeq 10^{-10}\,,     && T_{\rm re} \simeq 10^{13} \ \rm GeV\,, & & \\
    &\mbox{Modular model:} && \theta_{\rm in} = 0.01\,,         && V_0 \simeq 10^{-14}\,,            && H_p \simeq 10^{-10}\,,
    && \Gamma \simeq 10^{-11}\,,     && T_{\rm re} \simeq 10^{12} \ \rm GeV\,, & & \\
    &\mbox{Hyperbolic model:}     && \theta_{\rm in} = 0.01\,, && V_0 \simeq  10^{-14}\,,   && H_p \simeq 10^{-9}\,,
    && \Gamma \simeq 10^{-10}\,,     && T_{\rm re} \simeq 10^{13} \ \rm GeV\,.   
\end{aligned}
\end{equation}
We only have extremal bounds on the reheating temperature, where the lower one is set by Big Bang Nuclesynthesis (BBN) around $1$ MeV while upper one arises from the instantaneous reheating scenario which, for the universal class of single field models \cite{German:2022sjd}, is of order $10^{15}$ GeV. Then, for us to be able to neglect the reheating contribution to the post-inflationary dynamics of our models, we find that the perturbative reheating process can occur, at most, at $\sim10^{12}-10^{13} $ GeV. For reference, the Starobinsky model \cite{Starobinsky:1980te} predicts a perturbative reheating temperature, via the scalaron decay, of order $10^{9}$ GeV \cite{Gorbunov:2010bn}.

The estimates in Eq.~\eqref{Tre} should be regarded as a constraint on the possible perturbative reheating scenarios capable of being in agreement with the (post-)inflationary dynamics presented in this paper. Nonetheless, as the reheating window is very broad, one can certainly fit in a much wider class of reheating scenarios, which could be safely neglected three $e$-folds after the end of inflation. We could, in principle, take these constraints further, as shown in \cite{Cook:2015vqa, Garcia:2023tkk}, and study how considering reheating details can contribute to tighter bounds on the inflationary observables, leading to more robust predictions.

In addition to the perturbative reheating, oscillations of the classical field may lose their energy non-perturbatively, in the process of parametric resonance as described in \cite{Kofman:1994rk,Kofman:1997yn}.  This process may occur much faster, within $\mathcal{O}(10)$ oscillations. It would be interesting to check whether such a process may occur in the models under investigation. However,  the main part of the amplification of adiabatic perturbations, the formation of a plateau shown in Figs. \ref{dedekindplateau1} and \ref{PSJpot}, happens very early, within the first one to three oscillations. We do not expect any reheating-related effects at this early stage of the process.

However, we assume that eventually, after a long stage of oscillations, the fields $\vp$ and $\theta$ will decay, and the products of their decay will thermalize. After that, initial isocurvature perturbations disappear. The subject of our investigation was to check whether the isocurvature perturbations may affect the amplitude of curvature perturbations at the early stages of post-inflationary oscillations, and well before thermalization.

\section{Conclusions}\label{Discussion and conclusions}

In this work, we have studied two-field inflationary models in hyperbolic field space, where the inflaton field $\varphi$ follows an $\alpha$-attractor potential, and a nearly massless axion $\theta$. Potentials of this type appear, in particular, in the recently developed models with $SL(2,\mathbb Z)$-invariant potentials \cite{Kallosh:2024ymt}, but in this paper, we have extended our analysis to other models with potentials having similar properties.

Following up on the analytical investigation in Ref.~\cite{Kallosh:2024whb}, we have numerically verified that, in such models, the field $\theta$ does not move during inflation and thus, its fluctuations $\delta \theta$ do not feed into curvature perturbations until inflation ends and the field trajectory bends, \textit{i.e.}~when the fields begin to move towards one of the minima of the potential.

In the models under consideration, the amplitude of perturbations $\delta\theta$ at the moment of their production is very large -- as discussed in Eq.~\eqref{ampth} --, and it remains large during the subsequent stages of inflation. That is why these perturbations may lead to a transfer of power from isocurvature to adiabatic modes during the post-inflationary evolution. We have shown that, in many cases, the sourcing of curvature perturbations saturates very soon, within one to three oscillations after the end of inflation, resulting in a new plateau of the curvature power spectrum with an enhanced amplitude.

This enhancement depends on the choice of the inflationary model and on the initial value of the axion field $\theta_{\rm in}$, which remains frozen during inflation. We found that in some cases, the final results are highly sensitive to $\theta_{\rm in}$, suggesting the possibility of significant non-Gaussianity.  This issue is model-dependent; we leave a further investigation of this possibility to future work.

Meanwhile, there are two general consequences of our results: The spectral index $n_{s}$ of the adiabatic perturbations after their enhancement remains the same as in the single-field $\alpha$-attractors, whereas the tensor-to-scalar ratio $r$ decreases by a factor inversely proportional to the enhancement of the curvature power spectra Eq.~\eqref{decreaser}.

Finally, we should mention a somewhat different scenario,  complementary to the one studied in the main part of this paper: there is a large family of models with $SL(2,\mathbb Z)$-invariant potentials based on $j$-functions where the axion field can be stabilized, which either renders the field trajectory orthogonal to the axion perturbations, or suppresses axion perturbations alltogether  \cite{Carrasco:2025rud}.   The cosmological consequences of such models coincide with the predictions of the single-field $\alpha$-attractors. As argued in the Appendix \ref{A comment on the stabilization of the modular models}, by combining such potentials with other  $SL(2,\mathbb Z)$-invariant potentials, one can achieve axion stabilization in a more general set of theories.

While set in the specific context of hyperbolic models with specific scalar potentials, we anticipate that the possibility of interesting post-inflationary evolution is more general, and could also apply to other multi-field models.  Key ingredients include a geodesic inflationary phase with light isocurvature modes, such that all dynamics following the end of inflation can lead to an interesting sourcing scenario, transitioning from isocurvature to adiabatic modes.  In such cases, including the post-inflationary evolution is not optional but mandatory, in order to correctly extract inflationary predictions. It will be interesting to see how this effect can be applied to other scenarios.

\vspace{2mm}
\noindent{\bf Acknowledgments.} We are grateful to A.~Ach\'{u}carro, P.~Christodoulidis, M.~Braglia, E.~Copeland, L.~Iacconi, M.~Fasiello, D.~L\"{u}st, S.~Mishra and D.~Wands for their helpful comments. RGQ thanks CONAHCyT (SECIHTI) and RUG for funding and support. DR would like to thank the SITP for its hospitality and stimulating atmosphere during the initial phases of the research presented in this paper.  
RK and AL are supported by SITP and by the US National Science Foundation Grant PHY-2310429. JJMC is supported in part by the DOE under contract DE-SC0015910.

\appendix

\section{Modular inflation with the $j$-function}\label{J function}

\subsection{Inflationary and post-inflationary evolution}

Previous work \cite{Kallosh:2024whb} on modular cosmology has focused on a different $SL(2,\mathbb Z)$-invariant model. In this work, the potential reads
\begin{equation}\label{jPotential}
    V(\tau)=V_{0}\left(1-\frac{(\ln \beta^2)}{\ln(|j(\tau)|^{2}+\beta^{2})}\right)\,,
\end{equation}
where the $j$-invariant is defined as
\begin{equation}
    j(\tau)=q^{-1}+\sum_{n=0}c_{n}q^{n}\, \quad \mbox{where} \quad q=e^{2\pi i \tau} \,,
\end{equation}
and we set $\beta\equiv j(i)=12^3  $. This potential is similar to the one in Eq.~\eqref{dedekindmodel}, and exhibits a double exponential suppression in the $\theta$ direction.

\begin{figure}[t!]
    \centering
       \begin{subfigure}{0.45\textwidth}
        \centering
        \includegraphics[width=1\linewidth]{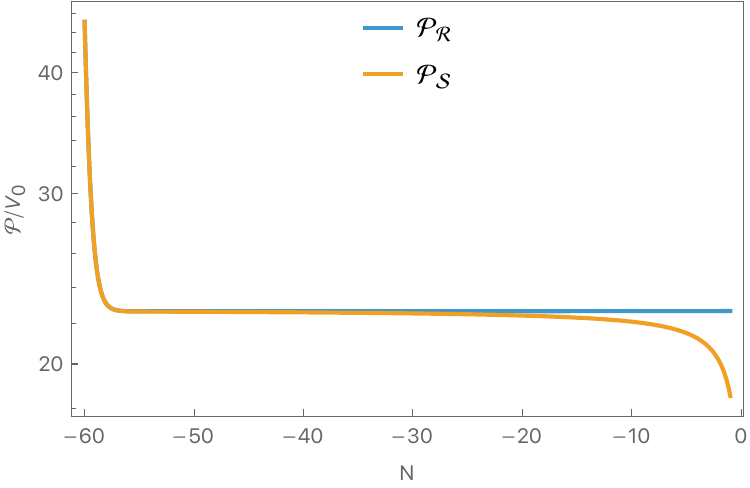}  
        \end{subfigure}
    \hfill
    \begin{subfigure}{0.45\textwidth}
        \centering
        \includegraphics[width=1\linewidth]{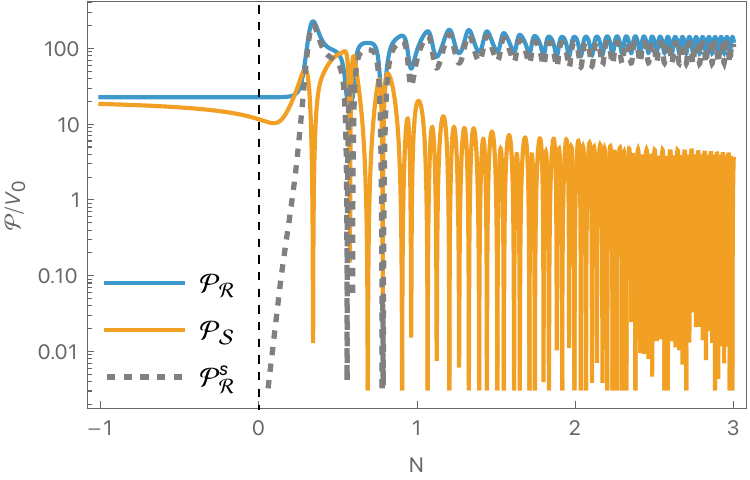} 
        \end{subfigure}
    \vfill
    \begin{subfigure}{0.45\textwidth}
        \centering
        \includegraphics[width=1\linewidth]{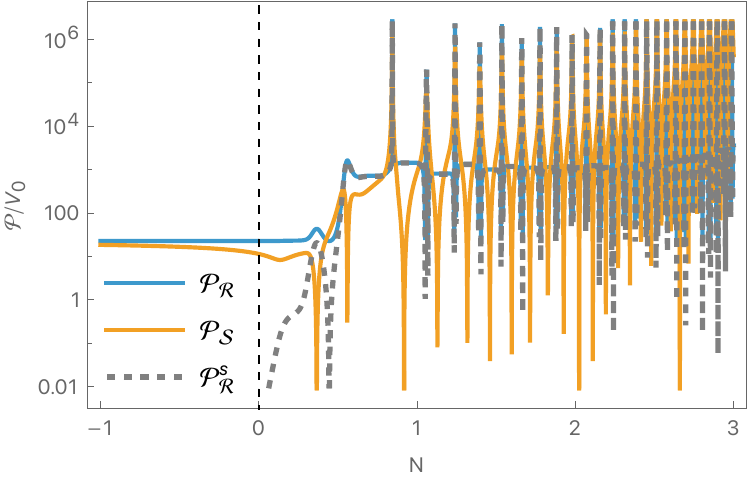} 
        \end{subfigure}\caption{\footnotesize{Plots of the curvature and isocurvature power spectra for the $j$-function model Eq.~\eqref{jPotential} during inflation (left panel) and after inflation (right panel). During inflation, the main component of the total curvature power comes from its vacuum contribution, while after the end of inflation -- when the sourcing of isocurvature into curvature perturbations is present -- the main contribution to the curvature power becomes its sourced component. We illustrate this for $\theta_{\rm in}=0.1$ (upper panel), where we do find a plateau within the first post-inflationary e-folds, and for $\theta_{\rm in}=0.3$ (lower panel), where the plateau is hidden behind the still growing isocurvature modes. This precise behavior may indicate that the plateau we see in the bottom panel is not stable if we evolve for a longer amount of $e$-folds. }}
    \label{PSJpot}
\end{figure}

Qualitatively, we thus expect to find similar results to those in the main text. Indeed, one can verify that, during inflation, the turning rate is extremely small, leading to a negligible sourcing of adiabatic perturbations from entropic perturbations, see the left upper panel of Fig.  \ref{PSJpot}. This confirms the results in \cite{Kallosh:2024whb} where, analytically, it was estimated that this sourcing would be suppressed due to the geodesic field space motion. Moreover, for $\theta_{\rm in}=0.1$, after the end of inflation, when the field space trajectory experiences significant turns, we again find an efficient enhancement of adiabatic perturbations, as can be seen from the upper right and lower panels of Fig.  \ref{PSJpot}.

Unlike the Dedekind modular model, the $j$-function model exhibits a key difference: only in the vicinity of $\theta_{\rm in}=0.1$ is the sourcing such that we find a plateau in the curvature power after a couple of $e$-folds after the end of inflation. Generally, we find that both the curvature and isocurvature perturbations continue to grow up to $N\sim 3$ and beyond. We illustrate this for $\theta_{\rm in}=0.3$ in the lower panel of  Fig. \ref{PSJpot}.

Both the $ j$- and $\eta_{D}$-functions modular inflationary models therefore lead to qualitatively similar results. Before the end of inflation, the $\theta$ direction experiences a double exponential suppression; thus, no turning in field space occurs. After the end of inflation we find that both models experience a significant, initial condition dependent, enhancement of curvature perturbations, where the main difference is that the Dedekind function always gives rise to a plateau, while for the $j$ for some initial values of $\theta$, the exponential amplification continues much longer, taking us beyond the limits of applicability of our methods. We attribute this to the nature of the Minkowski minima, which are of quartic nature for the potential in Eq.~\eqref{jPotential} while they are of quadratic nature for the two models studied in the main text.

\subsection{A comment on the stabilization of modular models}\label{A comment on the stabilization of the modular models}
The situation becomes very different if one considers a more general $SL(2,\mathbb Z)$-invariant potentials depending on the $j$ function, leading to axion stabilization \cite{Carrasco:2025rud}. One of such potentials is 
\be
V =V_0\Bigg (1-{\ln \beta^{2}
\over  \ln\Big[|j(\tau)|^2+A\,
| j(\tau) -\overline {j(\tau)}|^{2} + \beta^{2}\Big]}\Bigg ) \ .
\label{stab1}\ee
We will refer to the term with coefficient $A$ (which we assume to be positive) as the stabilization term, for reasons that will become apparent. This potential is $SL(2,\mathbb{Z})$ invariant because $ j(\tau)$ and $\overline {j(\tau)}$ are $SL(2,\mathbb{Z})$ invariants. This potential coincides with \ref{jPotential} for $A = 0$, but also for any $A$ along the directions $\theta = n/2$ where $n$ is an integer.

At large $\vp$, this potential is given by 
 \be
  V_\beta^{j-\bar j}  = V_0 \left( 1 - \frac{\ln \beta}{2\pi} e^{- \sqrt{2\over 3\alpha} \varphi} + \frac{\ln \beta}{8 \pi^2} e^{-2\sqrt{2\over 3\alpha} \varphi} \ln(1 + 4 A  \sin^2(2 \pi \theta) )+ \ldots \right) \,.  
 \label{stab2}
 \ee
The last term in this expression leads to the axion stabilization along the axion valleys with $\theta = n/2$.

The dynamical mechanism of stabilization consists of two parts, each of which plays an important role. First of all, one may consider various initial conditions for the fields $\vp$ and $\theta$ and find the subsequent evolution of these fields. In the absence of stabilization, the field $\theta$, even if it was moving, loses its kinetic energy within a few $e$-folds and stops its motion until the end of inflation. Meanwhile, stabilization forces the field to move towards the axion valleys with $\theta = n/2$.

In the slow-roll regime with $A\ll1$, the relative speed of motion of the fields is given by  \cite{Carrasco:2025rud}
\be
{\dot\theta \over \dot\vp} =    4 A\sqrt{2\over 3\alpha}\,  e^{ \sqrt{2\over 3 \alpha}\vp}\     \sin 4\pi \theta 
\label{stab3} \ .
\ee
This means that if inflation begins at a sufficiently large $\vp$, the fields always have enough time to approach one of the axion valleys  $\theta = n/2$, and they stay there until the end of inflation.

We emphasize that this conclusion is true even if the stabilizing factor $A$ is extremely small, because the exponentially large factor $e^{ \sqrt{2\over 3 \alpha}\vp}$  can overcome the smallness of $A$ if inflation begins at sufficiently large $\vp$.

The resulting scenario is different from the one studied in the present paper.  Suppose, for example, that the field $\theta$ initially was somewhere in the broad range $1/4< \theta < 3/4$. Then, during the early stages of inflation, it moves extremely closely to  $\theta = 0.5$. For $\theta = 0.5$,  the field continues moving along the same trajectory towards the minimum of the potential at $\theta = 0.5$, even after inflation, and the only source of the curvature perturbations is the perturbations of the field $\vp$. In that case, the cosmological predictions should coincide with the standard predictions of the single-field $\alpha$-attractors.

A stronger statement can be made for $ A \gtrsim 0.02$. In that case, the axion field falls to the axion valley even faster, and the axion mass in this valley is much greater than the Hubble constant. In this case, isocurvature perturbations are not generated at all, and, just as in the previous case,  the cosmological predictions coincide with the standard predictions of the single-field $\alpha$-attractors   \cite{Carrasco:2025rud}.

Now we are ready to bring it all together. Consider a combination of the Dedekind modular potential Eq.~\eqref{dedekindmodel} and the stabilized $j$-function potential
\be\label{dedJJ}
    V_{DJ}=V_{D}\left(\frac{1-g_0/g(\varphi,\theta)}{1+g_0/g(\varphi,\theta)}\right)  +V_J\Big (1-{\ln \beta^{2}
\over  \ln\Big[|j(\tau)|^2+A\,
| j(\tau) -\overline {j(\tau)}|^{2} + \beta^{2}\Big]}\Big ) \ ,
\ee
where $g(\varphi,\theta)$ is defined in Eq.~\eqref{fg}. At $A = 0$, both of these potentials are extremely flat in the axion direction, as shown in \cite{Kallosh:2024whb} and in our equation \rf{largephipot}. For $A > 0$, the potential $V_{DJ}$ has the axion valleys at $\theta = n/2$.

Following the same analysis as in   \cite{Carrasco:2025rud}, one can show that even if the parameter $A$ or the potential $V_{j}$ is small, all inflationary trajectories beginning at sufficiently large $\vp$  but different $\theta$ will rapidly converge at the axion valleys and effectively merge there into the trajectories with $\theta = n/2$.  As an example, consider the field trajectory along the valley with $\theta = 0.5$. It will not bend in the $\theta$ direction, not only during inflation, but also after inflation, on its way to the minimum of the potential at $\theta = 0.5$. This alone may already be sufficient to prevent the isocurvature perturbations from affecting the curvature perturbations.

Moreover, if the parameters $A$ and $V_{J}$ are sufficiently large, the axion mass along the axion valleys will be greater than the Hubble constant. In that case, there will be no isocurvature perturbations, and therefore the cosmological predictions of inflation along the axion valley with $\theta = 0.5$ will coincide with the main predictions of the single field $\alpha$-attractors.

\

\bibliographystyle{JHEP}
\bibliography{biblio}

\providecommand{\href}[2]{#2}\begingroup\raggedright\begin{thebibliography}{10}

\bibitem{Guth:1980zm}
A.~H. Guth, \emph{{The Inflationary Universe: A Possible Solution to the Horizon and Flatness Problems}}, \href{https://doi.org/10.1103/PhysRevD.23.347}{\emph{Phys. Rev. D} {\bfseries 23} (1981) 347}.

\bibitem{Linde:1984ir}
A.~D. Linde, \emph{{The Inflationary Universe}}, \href{https://doi.org/10.1088/0034-4885/47/8/002}{\emph{Rept. Prog. Phys.} {\bfseries 47} (1984) 925}.

\bibitem{Lyth:1998xn}
D.~H. Lyth and A.~Riotto, \emph{{Particle physics models of inflation and the cosmological density perturbation}}, \href{https://doi.org/10.1016/S0370-1573(98)00128-8}{\emph{Phys. Rept.} {\bfseries 314} (1999) 1} [\href{https://arxiv.org/abs/hep-ph/9807278}{{\ttfamily hep-ph/9807278}}].

\bibitem{Linde:1983gd}
A.~D. Linde, \emph{{Chaotic Inflation}}, \href{https://doi.org/10.1016/0370-2693(83)90837-7}{\emph{Phys. Lett. B} {\bfseries 129} (1983) 177}.

\bibitem{Linde:1993cn}
A.~D. Linde, \emph{{Hybrid inflation}}, \href{https://doi.org/10.1103/PhysRevD.49.748}{\emph{Phys. Rev. D} {\bfseries 49} (1994) 748} [\href{https://arxiv.org/abs/astro-ph/9307002}{{\ttfamily astro-ph/9307002}}].

\bibitem{Berera:2003yyp}
A.~Berera, \emph{{Warm inflation solution to the eta problem}}, \href{https://doi.org/10.22323/1.010.0069}{\emph{PoS} {\bfseries AHEP2003} (2003) 069} [\href{https://arxiv.org/abs/hep-ph/0401139}{{\ttfamily hep-ph/0401139}}].

\bibitem{Easson:2009kk}
D.~A. Easson and R.~Gregory, \emph{{Circumventing the eta problem}}, \href{https://doi.org/10.1103/PhysRevD.80.083518}{\emph{Phys. Rev. D} {\bfseries 80} (2009) 083518} [\href{https://arxiv.org/abs/0902.1798}{{\ttfamily 0902.1798}}].

\bibitem{Ashoorioon:2011aa}
A.~Ashoorioon, U.~Danielsson and M.~M. Sheikh-Jabbari, \emph{{1/N resolution to inflationary \ensuremath{\eta}-problem}}, \href{https://doi.org/10.1016/j.physletb.2012.06.034}{\emph{Phys. Lett. B} {\bfseries 713} (2012) 353} [\href{https://arxiv.org/abs/1112.2272}{{\ttfamily 1112.2272}}].

\bibitem{Freese:1990rb}
K.~Freese, J.~A. Frieman and A.~V. Olinto, \emph{{Natural inflation with pseudo - Nambu-Goldstone bosons}}, \href{https://doi.org/10.1103/PhysRevLett.65.3233}{\emph{Phys. Rev. Lett.} {\bfseries 65} (1990) 3233}.

\bibitem{Adams:1992bn}
F.~C. Adams, J.~R. Bond, K.~Freese, J.~A. Frieman and A.~V. Olinto, \emph{{Natural inflation: Particle physics models, power law spectra for large scale structure, and constraints from COBE}}, \href{https://doi.org/10.1103/PhysRevD.47.426}{\emph{Phys. Rev. D} {\bfseries 47} (1993) 426} [\href{https://arxiv.org/abs/hep-ph/9207245}{{\ttfamily hep-ph/9207245}}].

\bibitem{Kallosh:2024ymt}
R.~Kallosh and A.~Linde, \emph{{SL(2,\ensuremath{\mathbb{Z}}) cosmological attractors}}, \href{https://doi.org/10.1088/1475-7516/2025/04/045}{\emph{JCAP} {\bfseries 04} (2025) 045} [\href{https://arxiv.org/abs/2408.05203}{{\ttfamily 2408.05203}}].

\bibitem{Kallosh:2024pat}
R.~Kallosh and A.~Linde, \emph{{Landscape of Modular Cosmology}},  \href{https://arxiv.org/abs/2411.07552}{{\ttfamily 2411.07552}}.

\bibitem{Kallosh:2024whb}
R.~Kallosh and A.~Linde, \emph{{Double Exponents in $SL(2,\mathbb{Z})$ Cosmology}},  \href{https://arxiv.org/abs/2412.19324}{{\ttfamily 2412.19324}}.

\bibitem{Ding:2024euc}
G.-J. Ding, S.-Y. Jiang, Y.~Xu and W.~Zhao, \emph{{Modular invariant inflation, reheating and leptogenesis}},  \href{https://arxiv.org/abs/2411.18603}{{\ttfamily 2411.18603}}.

\bibitem{Aoki:2024ixq}
S.~Aoki and H.~Otsuka, \emph{{Inflationary constraints on the moduli-dependent species scale in modular invariant theories}},  \href{https://arxiv.org/abs/2411.08467}{{\ttfamily 2411.08467}}.

\bibitem{Carrasco:2025rud}
J.~J. Carrasco, R.~Kallosh, A.~Linde and D.~Roest, \emph{{Axion Stabilization in Modular Cosmology}},  \href{https://arxiv.org/abs/2503.14904}{{\ttfamily 2503.14904}}.

\bibitem{Aoki:2025wld}
S.~Aoki, H.~Otsuka and R.~Yanagita, \emph{{Higgs-Modular Inflation}},  \href{https://arxiv.org/abs/2504.01622}{{\ttfamily 2504.01622}}.

\bibitem{Kallosh:2013hoa}
R.~Kallosh and A.~Linde, \emph{{Universality Class in Conformal Inflation}}, \href{https://doi.org/10.1088/1475-7516/2013/07/002}{\emph{JCAP} {\bfseries 07} (2013) 002} [\href{https://arxiv.org/abs/1306.5220}{{\ttfamily 1306.5220}}].

\bibitem{Roest:2013fha}
D.~Roest, \emph{{Universality classes of inflation}}, \href{https://doi.org/10.1088/1475-7516/2014/01/007}{\emph{JCAP} {\bfseries 01} (2014) 007} [\href{https://arxiv.org/abs/1309.1285}{{\ttfamily 1309.1285}}].

\bibitem{Ferrara:2013rsa}
S.~Ferrara, R.~Kallosh, A.~Linde and M.~Porrati, \emph{{Minimal Supergravity Models of Inflation}}, \href{https://doi.org/10.1103/PhysRevD.88.085038}{\emph{Phys. Rev. D} {\bfseries 88} (2013) 085038} [\href{https://arxiv.org/abs/1307.7696}{{\ttfamily 1307.7696}}].

\bibitem{Kallosh:2013maa}
R.~Kallosh and A.~Linde, \emph{{Non-minimal Inflationary Attractors}}, \href{https://doi.org/10.1088/1475-7516/2013/10/033}{\emph{JCAP} {\bfseries 10} (2013) 033} [\href{https://arxiv.org/abs/1307.7938}{{\ttfamily 1307.7938}}].

\bibitem{Kallosh:2013daa}
R.~Kallosh and A.~Linde, \emph{{Multi-field Conformal Cosmological Attractors}}, \href{https://doi.org/10.1088/1475-7516/2013/12/006}{\emph{JCAP} {\bfseries 12} (2013) 006} [\href{https://arxiv.org/abs/1309.2015}{{\ttfamily 1309.2015}}].

\bibitem{Kallosh:2013yoa}
R.~Kallosh, A.~Linde and D.~Roest, \emph{{Superconformal Inflationary $\alpha$-Attractors}}, \href{https://doi.org/10.1007/JHEP11(2013)198}{\emph{JHEP} {\bfseries 11} (2013) 198} [\href{https://arxiv.org/abs/1311.0472}{{\ttfamily 1311.0472}}].

\bibitem{Kallosh:2013tua}
R.~Kallosh, A.~Linde and D.~Roest, \emph{{Universal Attractor for Inflation at Strong Coupling}}, \href{https://doi.org/10.1103/PhysRevLett.112.011303}{\emph{Phys. Rev. Lett.} {\bfseries 112} (2014) 011303} [\href{https://arxiv.org/abs/1310.3950}{{\ttfamily 1310.3950}}].

\bibitem{Cecotti:2014ipa}
S.~Cecotti and R.~Kallosh, \emph{{Cosmological Attractor Models and Higher Curvature Supergravity}}, \href{https://doi.org/10.1007/JHEP05(2014)114}{\emph{JHEP} {\bfseries 05} (2014) 114} [\href{https://arxiv.org/abs/1403.2932}{{\ttfamily 1403.2932}}].

\bibitem{Kallosh:2014rga}
R.~Kallosh, A.~Linde and D.~Roest, \emph{{Large field inflation and double $\alpha$-attractors}}, \href{https://doi.org/10.1007/JHEP08(2014)052}{\emph{JHEP} {\bfseries 08} (2014) 052} [\href{https://arxiv.org/abs/1405.3646}{{\ttfamily 1405.3646}}].

\bibitem{Kallosh:2014laa}
R.~Kallosh, A.~Linde and D.~Roest, \emph{{The double attractor behavior of induced inflation}}, \href{https://doi.org/10.1007/JHEP09(2014)062}{\emph{JHEP} {\bfseries 09} (2014) 062} [\href{https://arxiv.org/abs/1407.4471}{{\ttfamily 1407.4471}}].

\bibitem{Galante:2014ifa}
M.~Galante, R.~Kallosh, A.~Linde and D.~Roest, \emph{{Unity of Cosmological Inflation Attractors}}, \href{https://doi.org/10.1103/PhysRevLett.114.141302}{\emph{Phys. Rev. Lett.} {\bfseries 114} (2015) 141302} [\href{https://arxiv.org/abs/1412.3797}{{\ttfamily 1412.3797}}].

\bibitem{Carrasco:2015pla}
J.~J.~M. Carrasco, R.~Kallosh and A.~Linde, \emph{{$\alpha $-Attractors: Planck, LHC and Dark Energy}}, \href{https://doi.org/10.1007/JHEP10(2015)147}{\emph{JHEP} {\bfseries 10} (2015) 147} [\href{https://arxiv.org/abs/1506.01708}{{\ttfamily 1506.01708}}].

\bibitem{Carrasco:2015rva}
J.~J.~M. Carrasco, R.~Kallosh and A.~Linde, \emph{{Cosmological Attractors and Initial Conditions for Inflation}}, \href{https://doi.org/10.1103/PhysRevD.92.063519}{\emph{Phys. Rev. D} {\bfseries 92} (2015) 063519} [\href{https://arxiv.org/abs/1506.00936}{{\ttfamily 1506.00936}}].

\bibitem{Carrasco:2015iij}
J.~J.~M. Carrasco, R.~Kallosh and A.~Linde, \emph{{Minimal supergravity inflation}}, \href{https://doi.org/10.1103/PhysRevD.93.061301}{\emph{Phys. Rev. D} {\bfseries 93} (2016) 061301} [\href{https://arxiv.org/abs/1512.00546}{{\ttfamily 1512.00546}}].

\bibitem{Kallosh:2016sej}
R.~Kallosh, A.~Linde, D.~Roest and T.~Wrase, \emph{{Sneutrino inflation with $\alpha$-attractors}}, \href{https://doi.org/10.1088/1475-7516/2016/11/046}{\emph{JCAP} {\bfseries 11} (2016) 046} [\href{https://arxiv.org/abs/1607.08854}{{\ttfamily 1607.08854}}].

\bibitem{Iacconi:2024hmg}
L.~Iacconi, M.~Bacchi, L.~F. Guimar{\~a}es and F.~T. Falciano, \emph{{Testing inflation on all scales: a case study with {\ensuremath{\alpha}}-attractors}}, \href{https://doi.org/10.1088/1475-7516/2025/06/004}{\emph{JCAP} {\bfseries 06} (2025) 004} [\href{https://arxiv.org/abs/2412.02544}{{\ttfamily 2412.02544}}].

\bibitem{German:2022sjd}
G.~Germ\'an, R.~G. Quaglia and A.~M.~M. Colorado, \emph{{Model independent bounds for the number of e-folds during the evolution of the universe}}, \href{https://doi.org/10.1088/1475-7516/2023/03/004}{\emph{JCAP} {\bfseries 03} (2023) 004} [\href{https://arxiv.org/abs/2212.03730}{{\ttfamily 2212.03730}}].

\bibitem{Iacconi:2023mnw}
L.~Iacconi, M.~Fasiello, J.~V{\"a}liviita and D.~Wands, \emph{{Novel CMB constraints on the {\ensuremath{\alpha}} parameter in alpha-attractor models}}, \href{https://doi.org/10.1088/1475-7516/2023/10/015}{\emph{JCAP} {\bfseries 10} (2023) 015} [\href{https://arxiv.org/abs/2306.00918}{{\ttfamily 2306.00918}}].

\bibitem{Carrasco:2015uma}
J.~J.~M. Carrasco, R.~Kallosh, A.~Linde and D.~Roest, \emph{{Hyperbolic geometry of cosmological attractors}}, \href{https://doi.org/10.1103/PhysRevD.92.041301}{\emph{Phys. Rev. D} {\bfseries 92} (2015) 041301} [\href{https://arxiv.org/abs/1504.05557}{{\ttfamily 1504.05557}}].

\bibitem{Kallosh:2025jsb}
R.~Kallosh and A.~Linde, \emph{{Attractors in Supergravity}},  \href{https://arxiv.org/abs/2503.13682}{{\ttfamily 2503.13682}}.

\bibitem{Iarygina:2020dwe}
O.~Iarygina, E.~I. Sfakianakis, D.-G. Wang and A.~Ach\'ucarro, \emph{{Multi-field inflation and preheating in asymmetric $\alpha$-attractors}},  \href{https://arxiv.org/abs/2005.00528}{{\ttfamily 2005.00528}}.

\bibitem{Achucarro:2017ing}
A.~Ach\'ucarro, R.~Kallosh, A.~Linde, D.-G. Wang and Y.~Welling, \emph{{Universality of multi-field $\alpha$-attractors}}, \href{https://doi.org/10.1088/1475-7516/2018/04/028}{\emph{JCAP} {\bfseries 04} (2018) 028} [\href{https://arxiv.org/abs/1711.09478}{{\ttfamily 1711.09478}}].

\bibitem{Wands:2007bd}
D.~Wands, \emph{{Multiple field inflation}}, \href{https://doi.org/10.1007/978-3-540-74353-8_8}{\emph{Lect. Notes Phys.} {\bfseries 738} (2008) 275} [\href{https://arxiv.org/abs/astro-ph/0702187}{{\ttfamily astro-ph/0702187}}].

\bibitem{Bartolo_2001}
N.~Bartolo, S.~Matarrese and A.~Riotto, \emph{Adiabatic and isocurvature perturbations from inflation: Power spectra and consistency relations}, \href{https://doi.org/10.1103/physrevd.64.123504}{\emph{Physical Review D} {\bfseries 64} (2001) }.

\bibitem{Pinol:2021nha}
L.~Pinol, \emph{{Multifield aspects in the early Universe : Inflation and Reheating}}, Ph.D. thesis, Sorbonne Universit\'e, Paris, Inst. Astrophys., 2021.

\bibitem{Iacconi:2021ltm}
L.~Iacconi, H.~Assadullahi, M.~Fasiello and D.~Wands, \emph{{Revisiting small-scale fluctuations in {\ensuremath{\alpha}}-attractor models of inflation}}, \href{https://doi.org/10.1088/1475-7516/2022/06/007}{\emph{JCAP} {\bfseries 06} (2022) 007} [\href{https://arxiv.org/abs/2112.05092}{{\ttfamily 2112.05092}}].

\bibitem{Fumagalli:2019noh}
J.~Fumagalli, S.~Garcia-Saenz, L.~Pinol, S.~Renaux-Petel and J.~Ronayne, \emph{{Hyper-Non-Gaussianities in Inflation with Strongly Nongeodesic Motion}}, \href{https://doi.org/10.1103/PhysRevLett.123.201302}{\emph{Phys. Rev. Lett.} {\bfseries 123} (2019) 201302} [\href{https://arxiv.org/abs/1902.03221}{{\ttfamily 1902.03221}}].

\bibitem{Christodoulidis:2019jsx}
P.~Christodoulidis, D.~Roest and E.~I. Sfakianakis, \emph{{Scaling attractors in multi-field inflation}}, \href{https://doi.org/10.1088/1475-7516/2019/12/059}{\emph{JCAP} {\bfseries 12} (2019) 059} [\href{https://arxiv.org/abs/1903.06116}{{\ttfamily 1903.06116}}].

\bibitem{Bjorkmo:2019fls}
T.~Bjorkmo, \emph{{Rapid-Turn Inflationary Attractors}}, \href{https://doi.org/10.1103/PhysRevLett.122.251301}{\emph{Phys. Rev. Lett.} {\bfseries 122} (2019) 251301} [\href{https://arxiv.org/abs/1902.10529}{{\ttfamily 1902.10529}}].

\bibitem{Achucarro:2019mea}
A.~Ach\'ucarro and Y.~Welling, \emph{{Orbital Inflation: inflating along an angular isometry of field space}},  \href{https://arxiv.org/abs/1907.02020}{{\ttfamily 1907.02020}}.

\bibitem{Achucarro:2019pux}
A.~Ach\'ucarro, E.~J. Copeland, O.~Iarygina, G.~A. Palma, D.-G. Wang and Y.~Welling, \emph{{Shift-symmetric orbital inflation: Single field or multifield?}}, \href{https://doi.org/10.1103/PhysRevD.102.021302}{\emph{Phys. Rev. D} {\bfseries 102} (2020) 021302} [\href{https://arxiv.org/abs/1901.03657}{{\ttfamily 1901.03657}}].

\bibitem{Iarygina:2023msy}
O.~Iarygina, M.~C.~D. Marsh and G.~Salinas, \emph{{Non-Gaussianity in rapid-turn multi-field inflation}}, \href{https://doi.org/10.1088/1475-7516/2024/03/014}{\emph{JCAP} {\bfseries 03} (2024) 014} [\href{https://arxiv.org/abs/2303.14156}{{\ttfamily 2303.14156}}].

\bibitem{Peterson:2010mv}
C.~M. Peterson and M.~Tegmark, \emph{{Non-Gaussianity in Two-Field Inflation}}, \href{https://doi.org/10.1103/PhysRevD.84.023520}{\emph{Phys. Rev. D} {\bfseries 84} (2011) 023520} [\href{https://arxiv.org/abs/1011.6675}{{\ttfamily 1011.6675}}].

\bibitem{Linde:1990flp}
A.~D. Linde, \emph{{Particle physics and inflationary cosmology}}, vol.~5. 1990, [\href{https://arxiv.org/abs/hep-th/0503203}{{\ttfamily hep-th/0503203}}].

\bibitem{McDonald:1999hd}
J.~McDonald, \emph{{Reheating temperature and inflaton mass bounds from thermalization after inflation}}, \href{https://doi.org/10.1103/PhysRevD.61.083513}{\emph{Phys. Rev. D} {\bfseries 61} (2000) 083513} [\href{https://arxiv.org/abs/hep-ph/9909467}{{\ttfamily hep-ph/9909467}}].

\bibitem{Kolb:2003ke}
E.~W. Kolb, A.~Notari and A.~Riotto, \emph{{On the Reheating Stage after Inflation}}, \href{https://doi.org/10.1103/PhysRevD.68.123505}{\emph{Phys. Rev. D} {\bfseries 68} (2003) 123505} [\href{https://arxiv.org/abs/hep-ph/0307241}{{\ttfamily hep-ph/0307241}}].

\bibitem{Starobinsky:1980te}
A.~A. Starobinsky, \emph{{A New Type of Isotropic Cosmological Models Without Singularity}}, \href{https://doi.org/10.1016/0370-2693(80)90670-X}{\emph{Phys. Lett. B} {\bfseries 91} (1980) 99}.

\bibitem{Gorbunov:2010bn}
D.~S. Gorbunov and A.~G. Panin, \emph{{Scalaron the mighty: producing dark matter and baryon asymmetry at reheating}}, \href{https://doi.org/10.1016/j.physletb.2011.04.067}{\emph{Phys. Lett. B} {\bfseries 700} (2011) 157} [\href{https://arxiv.org/abs/1009.2448}{{\ttfamily 1009.2448}}].

\bibitem{Cook:2015vqa}
J.~L. Cook, E.~Dimastrogiovanni, D.~A. Easson and L.~M. Krauss, \emph{{Reheating predictions in single field inflation}}, \href{https://doi.org/10.1088/1475-7516/2015/04/047}{\emph{JCAP} {\bfseries 04} (2015) 047} [\href{https://arxiv.org/abs/1502.04673}{{\ttfamily 1502.04673}}].

\bibitem{Garcia:2023tkk}
M.~A.~G. Garcia, G.~Germ\'an, R.~Gonzalez~Quaglia and A.~M.~M. Colorado, \emph{{Reheating constraints and consistency relations of the Starobinsky model and some of its generalizations}}, \href{https://doi.org/10.1088/1475-7516/2023/12/015}{\emph{JCAP} {\bfseries 12} (2023) 015} [\href{https://arxiv.org/abs/2306.15831}{{\ttfamily 2306.15831}}].

\bibitem{Kofman:1994rk}
L.~Kofman, A.~D. Linde and A.~A. Starobinsky, \emph{{Reheating after inflation}}, \href{https://doi.org/10.1103/PhysRevLett.73.3195}{\emph{Phys. Rev. Lett.} {\bfseries 73} (1994) 3195} [\href{https://arxiv.org/abs/hep-th/9405187}{{\ttfamily hep-th/9405187}}].

\bibitem{Kofman:1997yn}
L.~Kofman, A.~D. Linde and A.~A. Starobinsky, \emph{{Towards the theory of reheating after inflation}}, \href{https://doi.org/10.1103/PhysRevD.56.3258}{\emph{Phys. Rev. D} {\bfseries 56} (1997) 3258} [\href{https://arxiv.org/abs/hep-ph/9704452}{{\ttfamily hep-ph/9704452}}].

\end{thebibliography}\endgroup

\end{document}